\documentclass[a4paper,fleqn,usenatbib]{mnras}

\usepackage{newtxtext,newtxmath}

\usepackage[T1]{fontenc}
\usepackage{ae,aecompl}

\usepackage{graphicx}	
\usepackage{amsmath}	
\usepackage{amssymb}	
\usepackage{float}
\usepackage{rotating}
\usepackage{pdflscape}

\def\hi{H\,{\sc i}}
\def\htwo{H$_\mathrm{2}$}
\def\deg{$^{\circ}$}
\def\Halpha{H$\alpha$}
\def\SFRSD{$\Sigma_{\mathrm{SFR}}$}
\def\HISD{$\Sigma_{\mathrm{HI}}$}
\def\GASSD{$\Sigma_{\mathrm{gas}}$}
\def\HIISD{$\Sigma_{\mathrm{H_\mathrm{2}}}$}
\def\kms{km~s$^{-1}$}
\def\msun{M$_{\odot}$}
\def\msunppc{M$_{\odot}~\mathrm{pc}^{-2}$}
\def\qgas{$Q_\mathrm{gas}$}
\def\sgas{$S_\mathrm{gas}$}
\def\galex{\textit{GALEX}}
\def\spitzer{\textit{Spitzer}}
\def\micron{$\mu$m}
\def\SFRSDW3{$\Sigma_\mathrm{SFR_{W3}}$}
\def\SFRSDhybrid{$\Sigma_\mathrm{SFR_{FUV+24~{\mu m}}}$}

\title[Star Formation in M~33]{A Multi-Scale Study of Star Formation in Messier 33}

\author[E. C. Elson et al.]{
E. C. Elson$^{1,2}$
S. Z. Kam$^{3,4}$, 
L. Chemin$^{5}$,
C. Carignan$^{2,3,4}$,
T. H. Jarrett$^{2}$
\\
$^{1}$Department of Physics $\&$ Astronomy, University of the Western Cape, Cape Town 7535, South Africa\\
$^{2}$Department of Astronomy, University of Cape Town, Private Bag X3, Rondebosch 7701, South Africa\\
$^{3}$Laboratoire de Physique et de Chimie de l'Environnement, Observatoire d'Astrophysique de l'Universite Ouaga I Pr Joseph Ki-Zerbo (ODAUO), \\
03 BP 7021, Ouaga 03, Burkina Faso\\
$^{4}$Departement de Physique, Universite de Montreal, C. P. 6128, Succ. centre-ville, Montreal, QC, H3C 3J7, Canada\\
$^{5}$Centro de Astronom\'ia (CITEVA), Universidad de Antofagasta, Avenida Angamos 601 Antofagasta, Chile\\
}

\date{Accepted XXX. Received YYY; in original form ZZZ}

\pubyear{2018}

\begin{document}
\label{firstpage}
\pagerange{\pageref{firstpage}--\pageref{lastpage}}
\maketitle

\begin{abstract}
For the Local Group Scd galaxy M~33 this paper presents a multi-scale study of the relationship between the monochromatic star formation rate (SFR) estimator based on 12~\micron\ emission and the total SFR estimator based on a combination of far-ultraviolet and 24~\micron\ emission.  We show the 12~\micron\ emission to be a linear estimator of total SFR on spatial scales from 782~pc down to 49~pc, over almost four magnitudes in SFR.  These results therefore extend to sub-kpc length scales the analogous results from other studies for global length scales.  We use high-resolution \hi\ and $^{12}\mathrm{CO}(J=2-1)$ image sets from the literature to compare the star formation to the neutral gas.  For the full range of length scales we find well-defined power-law relationships between 12~\micron-derived SFR surface densities and neutral gas surface densities.  For the \htwo\ gas component almost all correlations are consistent with being linear.  No evidence is found for a breakdown in the star formation law at small length-scales in M~33 reported by other authors.  We show that the average star formation efficiency in M~33 is roughly $10^{-9}$~yr$^{-1}$ and that it remains constant down to giant molecular cloud length-scales.  Toomre and shear-based models of the star formation threshold are shown to inaccurately account for the star formation activity in the inner disc of M~33.  Finally, we clearly show that the \hi\ saturation limit of $\approx 9$~\msunppc\ reported in the literature for other galaxies is not an intrinsic property of M~33 - it is systematically introduced as an artefact of spatially smoothing the data.

\end{abstract}

\begin{keywords}
galaxies: evolution -- galaxies: ISM -- galaxies: kinematics and dynamics
\end{keywords}

\section{Introduction}
At a distance of 0.84~Mpc \citep{Kam_2015}, Messier 33 (M~33) has been used by many investigators to carry out resolved studies of the processes that drive its star formation activity.   The intermediate inclination of the disc of M~33 ($\approx 54^{\circ}$, \citealt{Kam_2015}) makes it a particularly useful galaxy for which to carry out dynamical studies as well as studies of the distributions of gas and star formation, and their links to one another.  

In order to study the processes by which neutral hydrogen within galaxies is converted into stars, a reliable star formation tracer is required.  {Since the launch of the Galaxy Evolution Explorer (GALEX, \citealt{Martin_2005a}), far-ultraviolet (FUV) observations of galaxies have served as one of the most direct tracers of the recent (10 - 100~Myr) star formation rate (SFR).  FUV is regularly used as a star formation tracer as it directly traces the photospheric emission from young, massive stars over short timescales.. However, the FUV emission is highly susceptible to attenuation from interstellar dust.  This problem is often dealt with by using integrated far-infrared (FIR) emission to trace the dust that is heated by ultraviolet and optical photons (\citealt{Calzetti_2007, Rieke_2009}, and references therein).   In recent years it has become standard practice to combine the FUV and \emph{Spitzer} 24~$\mu$m emission of galaxies to generate a hybrid measure of their total SFR \citep{Calzetti_2007, THINGS_Leroy}.   Dust emission at 24~$\mu$m  is known to peak around regions of active star formation.  However, a more diffuse 24~$\mu$m component extends throughout galaxies (between the star forming regions typically traced by HII regions).  

For cases in which FUV and FIR observations of a galaxy are not both available, a monochromatic SFR tracer can be used.  The W3 band of the Widefield Infrared Survey Explorer (WISE, \citealt{WISE_Wright}) has been shown to serve as a good SFR estimator.  The band extends from 7.5 to 16.5~$\mu$m, centred at about 11.6~$\mu$m and covers a complex variety of PAH features, nebular emission lines, and rotational lines of \htwo.  Recent studies (e.g., \citealt{Jarrett_2013, Cluver_2014, Cluver_2017, Brown_2017}) have shown SFRs derived from WISE~W3 imaging to be closely related to Balmer decrement corrected \Halpha\ luminosities.  However, such studies have focussed only on global star formation rates - none of them have probed the relations on smaller length scales.  In this work we capitalise on the proximity of M~33 to study the relationships between a W3-based monochromatic  SFR indicator and a hybrid SFR indicator based on FUV~+~24~$\mu$m emission on sub-kpc length scales ranging from 49~pc to 782~pc.  At all length scales, we demonstrate the clear existence of a linear relationship between the two SFR indicators.  

Having a reliable estimator of total SFR within a galaxy allows for studies of the quantitative relationships between the neutral gas content and SFR.  The Kennicutt-Schmidt law is the typical power law that is found to fit the correlation between gas and SFR surface densities in local galaxies on global scales \citep{Kennicutt_1989, Kennicutt_1998} and on sub-kpc scales larger than a few hundred parsecs(e.g., \citealt{Kennicutt_2007, THINGS_Bigiel, Kennicutt_Evans_2012}).  The slope of the power law has been found to vary from about 1 to 2 and seems to depend on the tracer of the gas surface density, $\Sigma_\mathrm{gas}$.  As recently summarised by \citet{Elmegreen_2018}, the power law index is consistently close to unity when considering CO surface densities (i.e., $\Sigma_\mathrm{gas}=\Sigma_\mathrm{H_2}$) in normal galaxies \citep{Wong_Blitz_2002, THINGS_Bigiel}.  For total gas surface densities (i.e., $\Sigma_\mathrm{gas}=\Sigma_\mathrm{HI+H_2}$) in galaxy discs, the power law index is typically found to be close to 1.4 \citep{Kennicutt_1998}.  Finally, power law slopes consistently close to 2 are found in dwarf irregular galaxies and in the outer parts (dominated by \hi) of spiral discs \citep{Elmegreen_Hunter_2015}.  \citet{Elmegreen_2018} has recently argued that thresholds are not needed to explain star formation correlations such as the various versions of the Kennicutt-Schmidt law.  Instead, he says, a single model of pervasive collapse can explain all of the correlations.  The various versions of the Kennicutt-Schmidt law are shown by \cite{Elmegreen_2018} to follow from the model if the selection effects of the observable used to define the dense gas fraction are taken into account.  In this work, we avoid introducing  such selection effects by  considering the full range of reliable gas surface densities presented by our data.  

The exact nature of the star formation law crucially affects the overall evolution of galaxies.  While the above-mentioned power law indices have been demonstrated on large length scales, any tight correlation between $\Sigma_\mathrm{SFR}$ and $\Sigma_\mathrm{gas}$ is generally treated as one that breaks down at smaller length scales (less that a few hundred parsecs).  The data sets we utilise in this work allow us to reliably study the SF law in M~33 down to physical resolutions as small as 49~pc.  We use our various map sets to also study the star formation efficiency (SFE) in M~33.  While global measures of SFE exist for many nearby galaxies, it is not well-known whether SFE changes significantly when approaching length scales of giant molecular clouds.  The proximity of M~33 makes it ideally suited to study SFE on very small length scales. 

A star formation threshold is used to understand which gas in a galaxy is actively forming stars.  The basic ingredients of most single-fluid models that consider only the gas properties are self-gravity, turbulence, and kinematics.  Quantitatively understanding how these various properties of the inter-stellar medium work together to either promote of inhibit star formation offers important insights into the workings of fundamental galaxy evolution processes.  In this work, we use the \hi\ and CO image sets of M~33 together with two of the best-known single-fluid star formation thresholds to assess their ability to correctly predict the presence of star formation in the galaxy.  

The layout of this paper is as follows.  In Section~\ref{Section_Data} we present and describe the basic properties of the image sets we use.  Section~\ref{SFRestimator} focuses on a multi-scale study of the relationship between a monochromatic SFR estimator based on \emph{WISE} W3 emission and a total SFR estimator based on a combination of far-ultraviolet and 24~\micron\ emission.   Having established the efficacy of the W3 SFR estimator, we use it in Section~\ref{SFlaws} to study the dependence of SFR and star formation efficiency on the neutral gas density in M~33.  Section~\ref{SF_thresholds} is based on a study of star formation thresholds in M~33.  In Section~\ref{saturation_limit}, we discuss the observed surface densities of our \hi\ image set in the context of \hi\ saturation limits reported by previous authors for other galaxies.  Finally, we present our conclusions in Section~\ref{summary}.

\section{Data}\label{Section_Data}
12~$\mu$m imaging of M~33 from the \textit{WISE} Enhanced Resolution Galaxy Atlas (WERGA, \citealt{Jarrett_2013}) is used in this work to calculate the monochromatic star formation rate of the galaxy.  M~33 is one of the galaxies in the WERGA sample for which super-resolution methods have been used to create a high spatial resolution set of images in each of the four \textit{WISE} bands.  The reader is referred to \citet{Masci_2009} and  \citet{WERGA1} for the details of the Maximum Correlation Method used to produce the high-resolution images.  The WERGA 12~$\mu$m map has a spatial resolution of $\sim~7.2$~arcsec and a pixel scale of 1~arcsec.   \citet{Brown_2017} present a relation between WISE 12~$\mu$m luminosity, $L_{W3}$, and Balmer decrement corrected $H\alpha$ luminosity, $L_\mathrm{H\alpha, Corr}$:
\begin{equation}
\log L_\mathrm{W3}=(40.79\pm 0.06) + (1.27\pm 0.04)\times (\log L_\mathrm{H\alpha, Corr} -40).
\end{equation}
This relation is used together with the relation between $L_\mathrm{H\alpha, Corr}$ and SFR from \citet{Kennicutt_2009} to convert the calibrated 12~$\mu$m map into a star formation rate surface density map, $\Sigma_\mathrm{SFR_{W3}}$, with units of \msun~yr$^{-1}$~kpc$^{-2}$ (Fig.~\ref{fig:maps}, top row).  This relation between monochromatic luminosity and SFR, as well as the relation presented below between far-ultraviolet and 24~\micron\ luminosities and SFR, is based on a \citet{Kroupa_2001} initial mass function.  The $\Sigma_\mathrm{SFR_{W3}}$ maps of M~33 yield a global SFR of 0.34~\msun~yr$^{-1}$.  The parameter uncertainties from the \citet{Brown_2017} relation were used to calculate upper and lower limits of  0.42~\msun~yr$^{-1}$ and 0.27~\msun~yr$^{-1}$, respectively, for the this W3-based global SFR.  

For visual comparison to our $\Sigma_\mathrm{SFR_{W3}}$ maps, the PACS 100~\micron\ map obtained as part of the Herschel M33 Extended Survey (HerM33es, \citealt{HerM33es}) is shown in the second row of Fig.~\ref{fig:maps}.  We have used this map to estimate a 100~\micron-based global SFR to compare to our 12~\micron-based global SFR.  To do this, we used the linear relation between 100~\micron\ emission and total SFR (H$\alpha$ + 24~\micron) for resolved star-forming regions in M~33, presented by \citet{Boquien_2010}:
\begin{equation}
\log\Sigma_\mathrm{SFR} = (0.998\pm 0.0244)\log\Sigma_\mathrm{100} - (35.8520\pm 0.8304),
\end{equation}
where $\Sigma_\mathrm{SFR}$ is in units of \msun~yr$^{-1}$~kpc$^{-2}$ and $\Sigma_{100}$ is the 100~\micron\ luminosity surface density in units of W~kpc$^{-2}$.  We quantified the Gaussian noise properties of the 100~\micron\ image and then applied a 1$\sigma$ flux cut.  The surviving pixels were converted to units of \msun~yr$^{-1}$~kpc$^{-2}$ using the relation above, and then summed to obtain the global SFR.  To incorporate the parameter uncertainties in the \citet{Boquien_2010} relation, we repeated the procedure 1000 times, each time using a uniformly distributed random number in the range -0.0244 to 0.0244 to represent the error in the first term of the relation, and a random number in the range -0.8304 to 0.8304 to represent the error in the second term.  Using all 1000 realisations of the $\Sigma_\mathrm{SFR}$ maps based on the \citet{Boquien_2010} relation, we obtain an estimate of the 100~\micron-based global SFR of $0.44\pm 0.10$~\msun~yr$^{-1}$.  This result is entirely consistent with our estimate of $0.34^{0.42}_{0.27}$~\msun~yr$^{-1}$ based on the 12~\micron\ imaging.

 \begin{figure*}
  \centering
  \includegraphics[width=2\columnwidth]{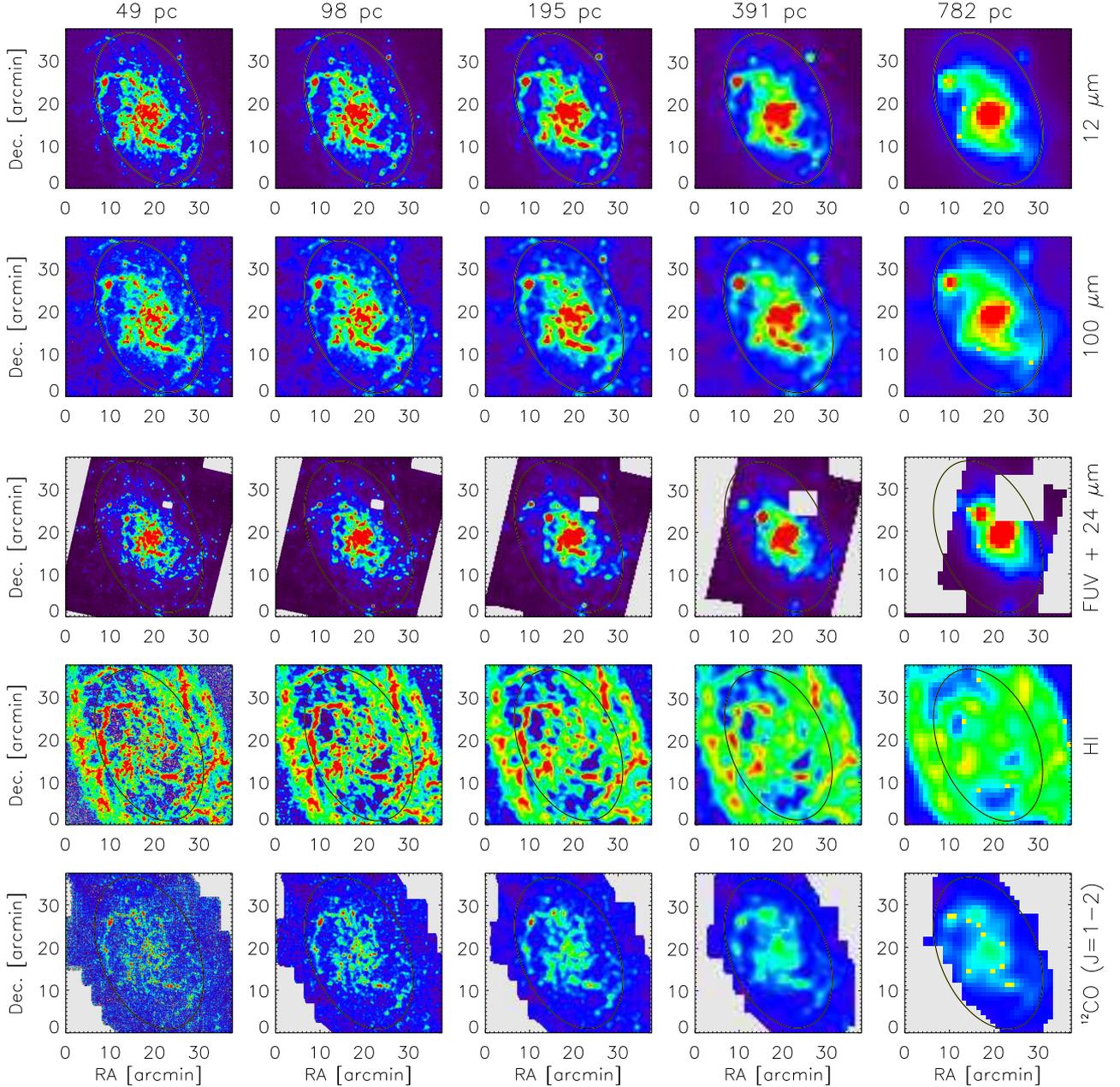}
  \caption{Representations of the various maps used in this work.  Shown from top to bottom are the $\Sigma_\mathrm{SFR_{W3}}$ map based on  WISE 12~$\mu$m imaging, the Herschel PACS 100~\micron\ map (shown here simply for visual comparison with the 12~$\mu$m map), the $\Sigma_\mathrm{SFR_{FUV+24_{\mu m}}}$ map based on GALEX FUV and \spitzer\ 24~$\mu$m imaging,  the \HISD\ map based on  JVLA imaging, and the \HIISD\ map based on $^{12}$CO($J$~=~2-1) IRAM imaging.  Shown from left to right are the maps at physical spatial resolutions of 49, 98, 195, 391, and 782 pc.  All maps in a row have their colours spanning a common SFR or mass surface density range.  From the top row to the bottom row, the colour scales span the following ranges: $-0.0025$ - 0.025~\msun~yr$^{-1}$~kpc$^{-2}$, $-0.002$ - 0.01~\msun~yr$^{-1}$~kpc$^{-2}$, $-0.0025$ - 0.4~\msun~yr$^{-1}$~kpc$^{-2}$, $-1$ - 17~\msun~pc$^{-2}$,  $-5$ - 20~\msun~pc$^{-2}$.    The  ellipse in each panel has a semi-major axis of length 18.6~arcmin, it delimits the inner portion of the galaxy for which various star formation thresholds are studied in Sec.~\ref{SF_thresholds}.  The increasing extent of the grey portion in  the $\Sigma_\mathrm{SFR_{FUV+24_{\mu m}}}$ and $^{12}$CO($J$~=~2-1) maps is due to blank pixels in the original images.}
  \label{fig:maps}
 \end{figure*}

In order to test the idea of using the 12~\micron\ imaging as a monochromatic SFR tracer in M~33, \textit{GALEX} far-ultraviolet (FUV) and \textit{Spitzer} SINGS 24~$\mu$m maps from the literature have been used to produce a hybrid tracer of the total SFR.  The  FUV map was obtained as part of the \textit{GALEX} Nearby Galaxies Survey \citep{NGS}.  Downloaded from the Barbara A. Mikulski Archive for Space Telescope,  it has a spatial resolution of 5.6~arcsec and a pixel scale of 1.5~arcsec.  The map was converted from \galex\ counts per second to magnitudes in the AB system, and then to units of MJy~ster$^{-1}$.  The dust map of \citet{Schlegel_1998} was used to obtain an estimate of $E(B-V)=0.0418$ for the reddening due to Galactic dust at the location of M~33.  The method of \citet{Wyder_2007} was then used to convert the reddening measure into an estimate of 0.32 magnitudes for the FUV extinction, which we applied to the FUV imaging.  The \spitzer\ 24~\micron\ map of M~33 used in this work is that from \citet{Tabatabaei_2007}, and was kindly provided by the authors of that study.  It is based on \spitzer\ Multiband imaging Photometer (MIPS, \citealt{MIPS}) observations of the galaxy performed on 9/10 January 2006.  \citet{Tabatabaei_2007} used the MIPS instrument team Data Analysis Tool to carry out the basic data reduction steps.  Extra steps were carried out to account for readout offset correction and array-averaged background subtraction.  The reader is referred to \citet{Tabatabaei_2007} for further details.  The final map with spatial resolution 6~arcsec, pixel scale 2.5~arcsec, and with units of MJy~ster$^{-1}$ is that shown in Fig.~2 of \citet{Tabatabaei_2007}.  With both the FUV and 24~\micron\ maps in units of MJy~ster$^{-1}$, the following prescription from \citet{THINGS_Leroy} was used to produce the total SFR map, $\Sigma_{\mathrm{SFR_{FUV+24_{\mu m}}}}$, in units of \msun~yr$^{-1}$~kpc$^{-2}$:
\begin{equation}
\Sigma_{\mathrm{SFR_{FUV+24_{\mu m}}}}=\left(8.1\times 10^{-2}I_\mathrm{FUV}+3.2\times 10^{-3}I_\mathrm{24~\mu m}\right)\cos i,
\end{equation}
where $I_\mathrm{FUV}$ and $I_\mathrm{24~\mu m}$ are the FUV and 24~\micron\ maps, respectively, and $i$ is the galaxy inclination (taken to be 53.9$^{\circ}$).  The \SFRSDhybrid\ maps of M~33 are shown in the third row of Fig.~\ref{fig:maps}.  The \SFRSDhybrid\ maps of M~33 yield a global SFR of  0.43~\msun~yr$^{-1}$.

We use the JVLA \hi\ data cube from \citet{Gratier_2010} to study the atomic neutral hydrogen in M~33.  The cube is based on archival JVLA B, C, and D array data taken as parts of projects AT206 and AT268 in 1997, 1998 and 2001.  It has a spatial resolution of $12~\times~11.6$~arcsec$^2$, a channel width of 1.27~\kms, and an rms noise of $\sigma_\mathrm{HI}=2.0$~mJy~beam$^{-1}$.  The reader is referred to  \citet{Gratier_2010} for the full details of the methods used to produce the cube.  In order to generate a variety of \hi\ data products from the cube, we fit a third-order Gauss-Hermite (GH$_3$) polynomial to every line profile.  The fitted profiles were used to create an \hi\ total intensity map (Fig.~\ref{fig:maps}, fourth row) by numerically integrating each GH$_3$ polynomial along the spectral axis of the cube.  

The $^{12}$CO($J$~=~2-1) total intensity map from \citet{Druard_2014} is used in this work to trace the molecular gas content of M~33.  The galaxy was observed in $^{12}$CO($J$~=~2-1) line emission with the HEterodyne Receiver Array (HERA, \citealt{Schuster_2004}) on the 30~m telescope of the Institut de RadioAstronomie Millimetrique (IRAM) on the Pico Veleta in southern Spain.  The $^{12}$CO($J$~=~2-1) total intensity map has a spatial resolution of 12~arcsec and a pixel scale of 3~arcsec.  \citet{Druard_2014} report a roughly constant value of 0.8 for the CO$\left({2-1\over 1-0}\right)$ intensity ratio, independent of radius.  An $N(\mathrm{H}_2)/I_{\mathrm{CO (1-0)}}= 4\times 10^{20}$~cm$^{-2}$/(K~\kms) conversion factor is used in this work to yield  H$_2$ mass surface density maps from the $^{12}$CO($J$~=~2-1) intensity map.  The \HIISD\ maps of M~33 are shown in the last row of Fig.~\ref{fig:maps}.

The main aim of this work is to study and compare the properties of the SFR activity and gas content of M~33 over a range of spatial scales, on a pixel-by-pixel basis.  For the five data sets mentioned above, the \hi\ and CO images have the lowest spatial resolution: 12~arcsec.  The corresponding physical resolution of 49~pc  therefore serves as the highest spatial resolution at which we can study the galaxy at multiple wavelengths.  Each of the  \SFRSDW3\ and \SFRSDhybrid\ maps were smoothed (using a Gaussian kernel of appropriate size) to a resolution of 12~arcsec and then re-sampled to have a pixel scale of 4~arcsec.  The \HIISD\ map was also re-sampled to have a pixel scale of 4~arcsec. Four more map sets were produced by smoothing the original images to spatial resolutions of 24, 48, 96, and 192~arcsec (corresponding to physical resolutions of ~ 98, 195, 391, and 782~pc for the assumed distance of 0.84~Mpc).  For each set of maps, the pixel scale was set to a third of the spatial resolution.  The various maps are shown from left to right in Fig.~\ref{fig:maps} in order of decreasing spatial resolution. 

\section{12~\micron\  emission as a SFR tracer}\label{SFRestimator}
One of the main aims of this work is to compare monochromatic SFRs derived from 12~\micron\ emission   to the surface density distribution of cold gas in M~33 in order to measure the star formation law.    Because the 12~\micron\ PAH  emission originates from the interstellar medium, it is an indirect tracer of the SFR.   A much more direct tracer of SFR is the ultraviolet photospheric emission of hot, massive stars with lifetimes of $\sim 100$~Myr \citep{Calzetti_2005, Salim_2007}.  This emission dominates the FUV band of \galex.  However, such a tracer potentially provides an incomplete view of the SF activity in a galaxy due to the presence of dust, which absorbs the FUV photons and re-emits them at infrared wavelengths.  \citet{Calzetti_2007} and \citet{Gonzalez_2006} demonstrated  the 24~\micron\ emission from a galaxy to be an accurate tracer of the dust-obscured SFR over timescales of $\sim 10$~Myr.  A hybrid SFR tracer generated by combining FUV and 24~\micron\ emission is generally regarded as an accurate tracer of the total SFR of a galaxy \citep{Calzetti_2005}.  In this work, we use the prescription given by \citet{THINGS_Leroy} to combine the FUV and 24~\micron\ maps of M~33 into a total SFR surface density map, \SFRSDhybrid.  Comparing this map to \SFRSDW3\ allows us to test the accuracy with which our 12~\micron-derived star formation rates approximate the total star formation rates.  


\citet{Brown_2017}  demonstrated the existence of a power law relation (index $1.27\pm 0.04$) between $WISE$ W3 (12~\micron) and Balmer decrement corrected \Halpha\ luminosities on \textit{global} length scales for a sample of 66 nearby star forming galaxies.  For galaxies in the combined SINGS and KINGFISH galaxy sample, \citet{Cluver_2017}  calibrated the \emph{global} $WISE$ W3 luminosities to SFRs derived using the total infrared luminosity.  The data had a best-fitting power law relation (index $1.124\pm 0.023$) over nearly 5 orders of magnitude in total infrared and 12~\micron\ luminosities.  In this section we aim to test whether these sorts of correlations seen on global length scales extend to more localised length scales in M~33

Figure~\ref{fig:SFRW3_SFRtotal} shows enlarged versions of our highest resolution (49~pc) \SFRSDW3\  and \SFRSDhybrid\ maps.  Both maps show only flux above 1$\sigma$.  The black contour is provided to facilitate direct visual comparison of the maps - it is at a level of \SFRSDW3~=~0.0058~\msun~yr$^{-1}$~kpc$^{-2}$ (corresponding to 10$\sigma$).  The two maps clearly have different morphologies, \textcolor{black}{and the difference becomes more pronounced at smaller length scales (i.e., higher resolutions)}.  The W3 map has at least two well-defined spiral arms that extend over most of the disc.  In fact, they extend approximately as far as the black ellipse shown in Fig.~\ref{fig:maps} (and reproduced in Fig.~\ref{fig:SFRW3_SFRtotal}), which is at a radius of $\sim 4.5$~kpc.  One arm extends to the north east while the other extends to the south west.  Compared to the W3 disc, the star-forming disc as traced by the FUV~+~24~\micron\ emission is smaller in radial extent.  It has a spiral morphology that is arguably less pronounced than that of the W3 disc.  Quite noticeable is the fact that much of the high-SFR activity ($\gtrsim 0.03$~~\msun~yr$^{-1}$~kpc$^{-2}$)  as seen in the FUV~+~24~\micron\ disc is concentrated near the centre of the galaxy.   Hence, while several galaxies in the local Universe have had their W3-based SFRs shown to be correlated with their total SFRs on \emph{global} length scales, M~33 clearly exemplifies the ways in which the structure of the two SFR tracers can vary on local length scales.

\begin{figure*}
	\includegraphics[width=1.8\columnwidth]{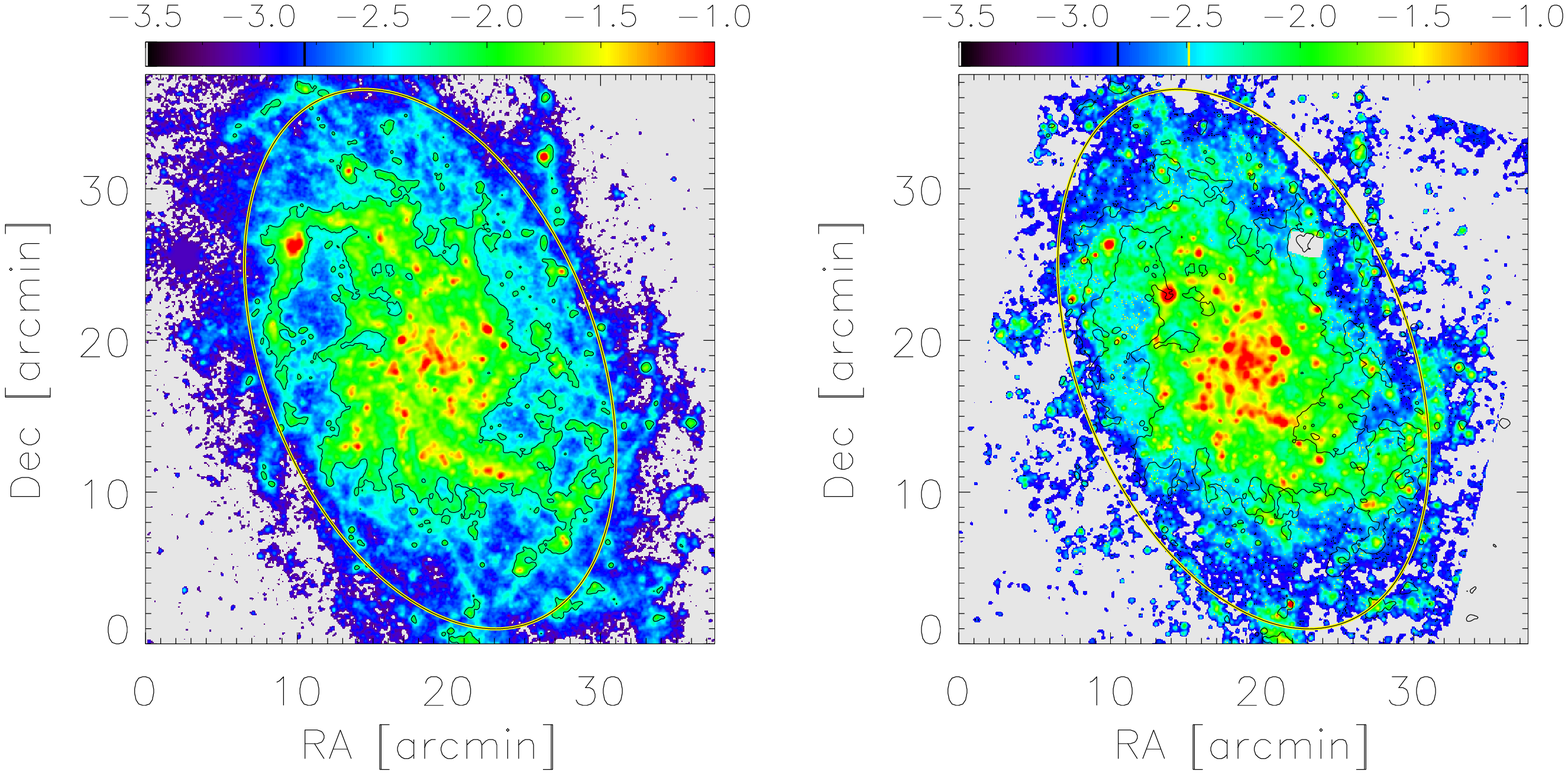}
    \caption{Enlarged versions of the 49~pc \SFRSDW3\ (left) and \SFRSDhybrid\ (right) maps shown in Fig.~\ref{fig:maps}.  Both maps show only flux above 1$\sigma$.  The same black contour shown in both maps is at a level \SFRSDW3~=~0.0058~\msun~yr$^{-1}$~kpc$^{-2}$ (corresponding to 10$\sigma$).  The colours in each map represent $\log_{10}\Sigma_\mathrm{SFR}~[\mathrm{M_{\odot}~yr^{-1}~kpc^{-2}}]$ as indicated by the colour bar.  The black ellipse in each panel is the same ellipse show in the panels of Fig.~\ref{fig:maps}, it has radius 18.6~arcmin ($\sim 4.5$~kpc).  \textcolor{black}{The morphology of M~33 clearly depends upon the wavelength at which the galaxy is observed. }}
    \label{fig:SFRW3_SFRtotal}
\end{figure*}

Despite M~33 having different W3 and FUV~+~24~\micron\ morphologies, we compare each pair of \SFRSDW3\ and \SFRSDhybrid\ maps corresponding to a particular physical resolution.  The maps are compared on a pixel-by-pixel basis.  The results are presented in two different ways in Fig.~\ref{fig:SFR_ratios}.  The top panels show $\log_{10}$(\SFRSDW3) as a function of $\log_{10}$(\SFRSDhybrid) for the maps with physical resolutions 49, 98, 195, 391, and 782 pc (left to right, with respective pixel sizes of 16, 32, 65, 130, and 260~pc). Clearly, for all spatial resolutions, there exists a an underlying linear relationship between \SFRSDW3\ and \SFRSDhybrid.  Furthermore, the relationship spans $\sim 2$ orders of magnitude in SFR above the 1$\sigma$ level.  However, going from low to high resolution does lead to a significant increase in the scatter of the relation.  To better demonstrate this, the bottom panels of Fig.~\ref{fig:SFR_ratios} show the distributions of \SFRSDW3/\SFRSDhybrid\ for the various resolutions.  In all cases, the distribution is roughly log-normal.  For a resolution of 49 pc, the 1$\sigma$ scatter about the mean is 0.32 dex.  The scatter steadily decreases to a 1$\sigma$ value of 0.19 dex when the resolution reaches 782~pc.  
\begin{figure*}
	\includegraphics[width=2.\columnwidth]{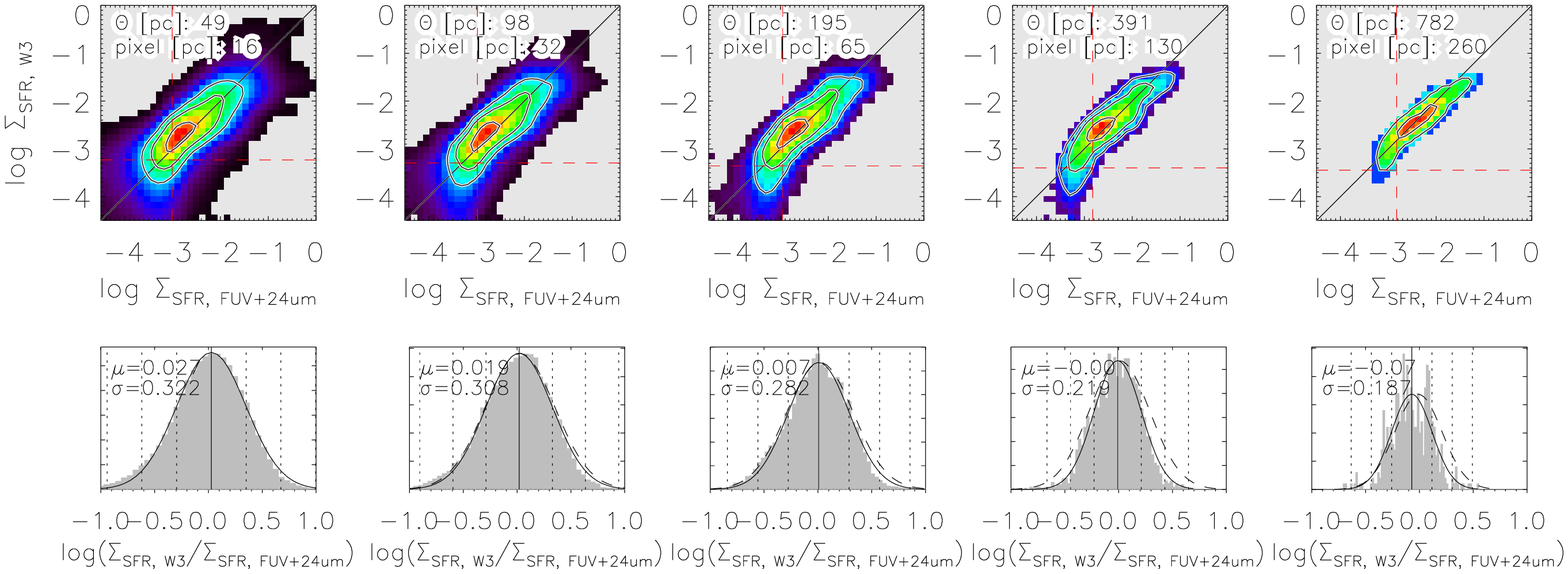}
    \caption{Relationships between \SFRSDW3\ and \SFRSDhybrid\ in M~33 for a range of spatial resolution scales.  Top row: $\log_{10}$(\SFRSDW3 [\msun\ yr$^{-1}$ kpc$^{-2}$]) as a function of $\log_{10}$(\SFRSDhybrid [\msun\ yr$^{-1}$ kpc$^{-2}$]).  Given in the top left of each panel is the spatial resolution of the map as well as the corresponding pixel scale (both in units of pc).  The solid black line in each panel is in no way fit to the data, it is the line $y=x$.  The black contours are provided simply to better highlight the general shape of the distribution, they occur at levels of 0.9, 0.95 and 0.99 times the maximum amplitude of a bin in the 2-D histogram.  The red-dashed lines in each panel represent the 1$\sigma$ levels of the SFR surface density.  \textcolor{black}{While the average relationship between \SFRSDW3\ and \SFRSDhybrid\ is very close to being linear, there clearly are high SFR regions in the galaxy at which the difference is as high as a factor $\sim 10$.  Figure \ref{fig:SFR_ratio_maps} shows the spatial distribution of the $\log_{10}$(\SFRSDW3/\SFRSDhybrid) maps}.  Bottom row: Distributions of $\log_{10}$(\SFRSDW3/\SFRSDhybrid) for the various resolutions.  At the top-left of each panel is shown the mean and standard deviation of a Gaussian fit to the distribution (solid black curve).  The vertical solid black line in each panel marks the mean of the Gaussian, the black-dotted vertical lines represent 1, 2, and 3 standard deviations either side of the mean.  For all but the left-most panel, the black-dashed curve represents the best-fit Gaussian from the panel immediately to the left (normalised to have the same peak amplitude).  The 12~\micron\ and FUV~+~24~\micron\ SFR estimators are clearly related to one another in a linear fashion.  However, scatter in the relation increases significantly with increasing spatial resolution.}
    \label{fig:SFR_ratios}
\end{figure*}

To better identify which regions of the galaxy have the highest scatter, we show in Fig.~\ref{fig:SFR_ratio_maps} spatial maps of $\log_{10}$(\SFRSDW3/\SFRSDhybrid) for various resolution.  As expected, the highest deviations between the two SFR measures can be attributed to the differing W3 and FUV~+~24~\micron\ morphologies of the galaxy.  Absolute differences between the two maps are largest along the prominent spiral arms seen in W3, as well as within  the inter-arm regions.  The largest differences are as large as a factor 10 in magnitude.  However, despite these localised regions where the differences between the two SFR maps are high, most of the galaxy has the two SFR estimators agreeing to within a factor 2.  This is the case not only over regions of very low SFR surface densities - most of the flux in these regions is above the 3$\sigma$ level.
\begin{figure*}
	\includegraphics[width=1.8\columnwidth]{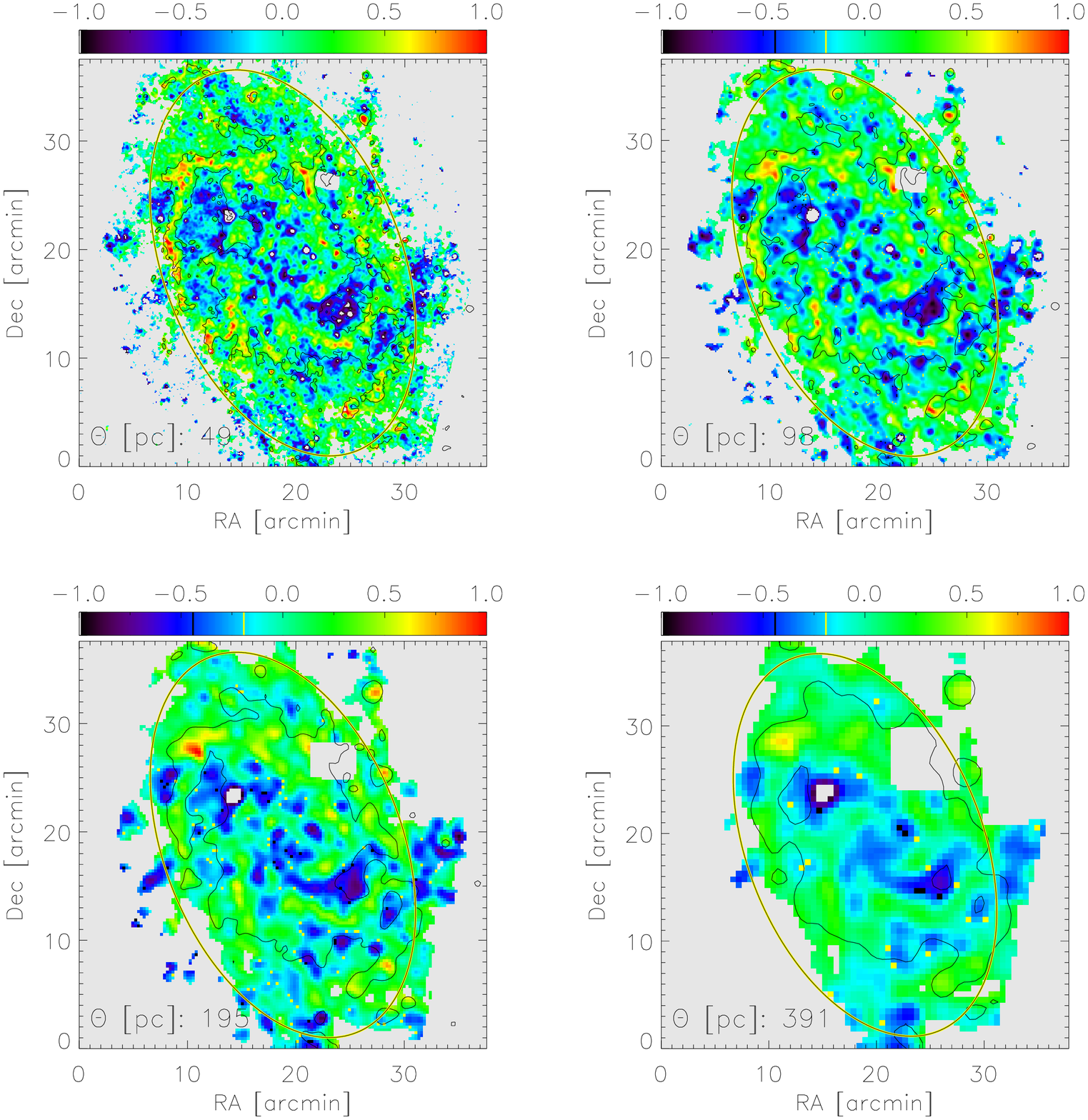}
    \caption{Maps of $\log_{10}$(\SFRSDW3/\SFRSDhybrid) for spatial resolutions 49, 98, 195, 391~pc.  The values associated with the colours in each maps are specified by the colour bar.  The solid black contours represent the 10$\sigma$ levels in the corresponding \SFRSDW3\ maps and are used to delimit the regions of highest SF activity.  Clearly, differences between W3-based SFR estimates and total SFR estimates are largest over regions of the galaxy where its morphologies in the bands differ most significantly.  The black ellipse in each panel is the same ellipse show in the panels of Fig.~\ref{fig:maps}, it has radius 18.6~arcmin ($\sim 4.5$~kpc).}
    \label{fig:SFR_ratio_maps}
\end{figure*}

The existence of a linear relationship between the monochromatic 12~\micron\ SFR estimator and the hybrid FUV+24~\micron\ SFR estimator is perhaps not surprising for the reasons mentioned by \citet{Cluver_2017}.  They point out that at $z=0$, the $WISE$~W3 band samples a variety of PAH emission features, the S(2) line of pure rotational \htwo, as well as at least two nebular emission lines ([Ne~$\mathrm{{\sc II}}$] and [Ne~$\mathrm{{\sc III}}$]).  PAH fractions are high in regions of active star formation, most likely due to them growing on dust grains in molecular clouds.  However, \citet{Cluver_2017} also mention that the $WISE$~W3 band is ``dominated by non-PAH continuum, coming from warm, large grains and stochastically heated grains''.

Our main conclusion in this section is that the 12~\micron\ emission in M~33, as observed in the W3 band of $WISE$, serves as a generally reliable linear estimator of the total star formation rate.  This is true over a range of sub-kpc length scales and over a large range of SFR surface densities.

\section{Star formation laws}\label{SFlaws}
Given that the 12~\micron-derived SFR is related to the total SFR in a generally linear way over a range of sub-kpc length scales, we can compare our various $\Sigma_\mathrm{SFR_\mathrm{W3}}$ maps to their corresponding \HISD\ and \HIISD\ maps on a pixel-by-pixel basis in order to carry out a multi-scale study of the star formation law in M~33.  This will be the focus of this section.

Figure~\ref{fig:SFRW3_laws} shows the distribution of $\left(\log_{10}\Sigma_\mathrm{gas},~\log_{10}\Sigma_\mathrm{SFR_{W3}}\right)$ pixel pairs for the cases in which only the \hi\ is considered for the neutral gas component ($\Sigma_\mathrm{gas}=\Sigma_\mathrm{HI}$, top row), the sum of the \hi\ and \htwo\ components is considered ($\Sigma_\mathrm{gas}=\Sigma_\mathrm{HI}+\Sigma_\mathrm{H_2}$, middle row), only the \htwo\ is considered ($\Sigma_\mathrm{gas}=\Sigma_\mathrm{H_2}$, bottom row).  Shown in columns from left to right in Fig.~\ref{fig:SFRW3_laws} are the relations between $\log_{10}\Sigma_\mathrm{SFR_{W3}}$ and $\log_{10}\Sigma_\mathrm{gas}$ at physical spatial resolutions of 49, 98, 195, 391, and 782~pc.  In order to robustly fit power laws to the two-dimensional data distributions shown in Fig.~\ref{fig:SFRW3_laws}, the 60, 70, 80, 90, and 95~percent percentiles of the number of points per two-dimensional bin were used to create a set of five contours (shown in black in each panel of Fig.~\ref{fig:SFRW3_laws}).  For each contour, the highest values of $\log_{10}$\GASSD\ and $\log_{10}\Sigma_\mathrm{SFR_{W3}}$ were used to define an ordered pair $\left(\log_{10}\Sigma_\mathrm{gas}^\mathrm{max}, \log_{10}\Sigma_\mathrm{SFR}^\mathrm{max}\right)$.  These are shown as red-filled circles in Fig.~\ref{fig:SFRW3_laws}.  A first-order polynomial was fit to each set of five $\left(\log_{10}\Sigma_\mathrm{gas}^\mathrm{max}, \log_{10}\Sigma_\mathrm{SFR}^\mathrm{max}\right)$ pairs in order to quantify the SF law in the form  $\Sigma_\mathrm{SFR}=A\Sigma_\mathrm{gas}^n$.  In each panel in Fig.~\ref{fig:SFRW3_laws}, the red-dashed horizontal and vertical lines represent the 3$\sigma$ levels of \SFRSDW3\ and $\Sigma_\mathrm{gas}$, respectively.

\begin{figure*}
	\includegraphics[width=2.15\columnwidth]{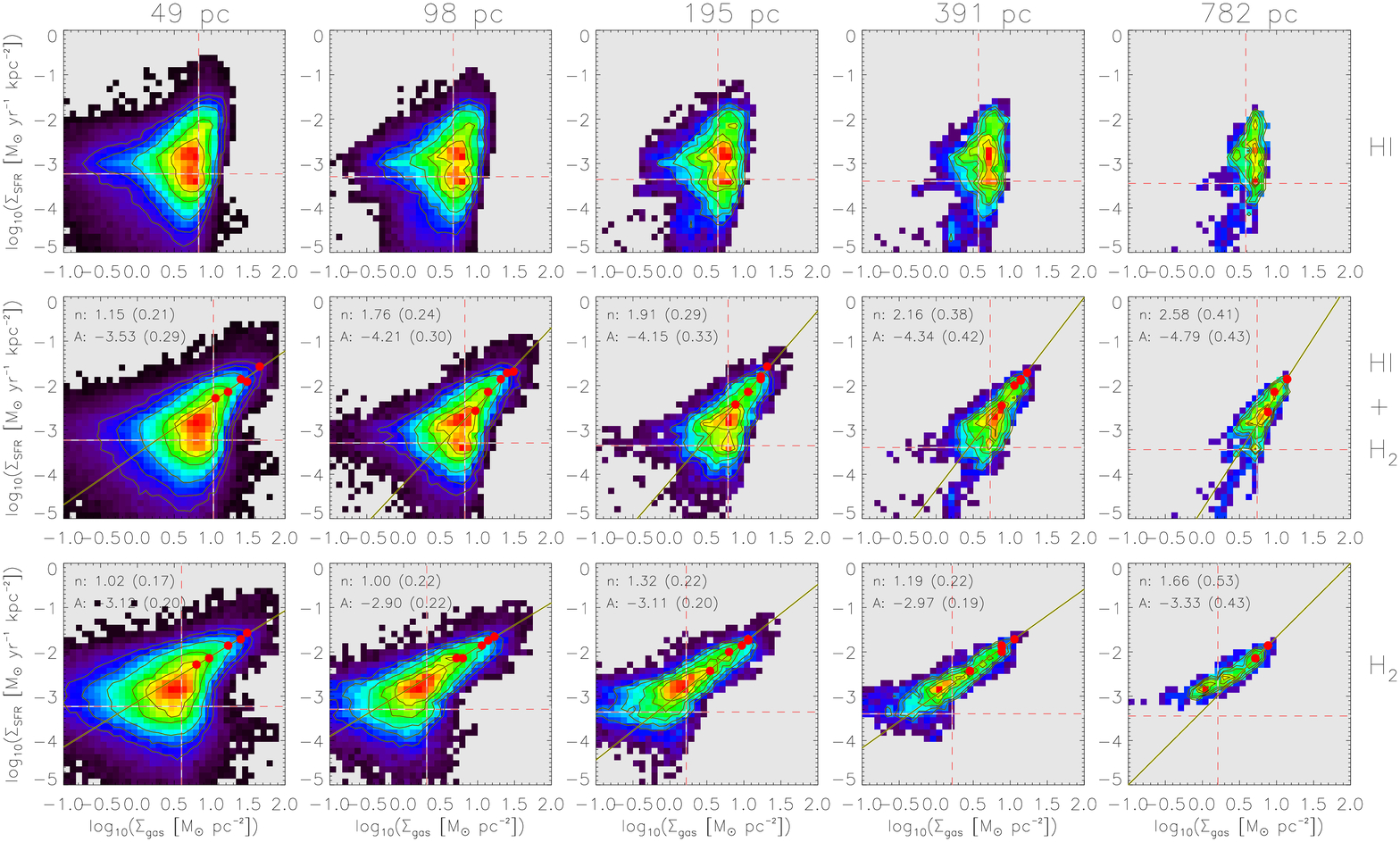}
	\caption{12~\micron-derived SFR surface density, $\Sigma_\mathrm{SFR_{W3}}$, shown as functions of neutral gas surface density, $\Sigma_\mathrm{gas}$.  From top to bottom, the rows represent the cases in which $\Sigma_\mathrm{gas}=\Sigma_\mathrm{HI},~\Sigma_\mathrm{HI+H_2},~\Sigma_\mathrm{H_2}$.  From left to right, the columns represent the surface density maps at spatial resolutions of 49, 98, 195, 391, 782~pc.  In each panel, the horizontal and vertical red-dashed lines represent, respectively, the 3$\sigma$ surface densities of the corresponding $\Sigma_\mathrm{gas}$ and $\Sigma_\mathrm{SFR_{W3}}$ maps.  The reliable portion of each 2-dimensional distribution is therefore the upper right quadrant.  The black contours in each panel mark the 60, 70, 80, 90, and 95~percent percentiles of the number of points per two-dimensional bin.  For each contour, the highest values of $\log_{10}$\GASSD\ and $\log_{10}\Sigma_\mathrm{SFR_{W3}}$ were used to define the red-filled circles.  The solid black line fit to the red-filled circles in each panel represents the best-fit Kennicutt-Schmidt star formation law in the form $\Sigma_\mathrm{SFR}=A\Sigma_\mathrm{gas}^n$.  For the middle and bottom rows, the best-fit parameters are shown in the top left corner of each panel.  \textcolor{black}{For all length scales, $\Sigma_\mathrm{SFR_{W3}}$ and $\Sigma_\mathrm{H_2}$ are clearly related by a power law with an index that is either fully consistent with a value of unity, or very close to being consistent.}}
    \label{fig:SFRW3_laws}
\end{figure*}

The most striking result from Fig.~\ref{fig:SFRW3_laws} comes from the comparison of W3-derived SFR and H$_2$ surface densities (bottom row of Fig.~\ref{fig:SFRW3_laws}).  For all length scales, the two quantities are clearly related by a power law with an index that is either fully consistent with a value of unity, or very close to being consistent.  All of the results are  also consistent with those of \citet{THINGS_Bigiel} who, for a subset of spiral galaxies from THINGS, measured $n=1.0\pm 0.2$.   However, \citet{THINGS_Bigiel} studied the star formation law at a fixed spatial resolution of 750~pc.  In this work, for M~33, we have shown similar star formation laws (at least in terms the the power law index) apply over a range of spatial resolutions from 49 to 782~pc.  The middle row of Fig.~\ref{fig:SFRW3_laws} clearly demonstrates the presence of power law relationships between $\Sigma_\mathrm{SFR_{W3}}$ and $\Sigma_\mathrm{HI+H_2}$.  However, at all spatial scales, the power law index is higher than that of the corresponding $\Sigma_\mathrm{H_2}$ case.  Finally, when considering only the \hi\ component of the neutral gas in M~33 (top row of Fig.~\ref{fig:SFRW3_laws}), \textcolor{black}{there is no well-defined relationship between  $\Sigma_\mathrm{SFR_{W3}}$ and $\Sigma_\mathrm{HI}$, presumably due to the manner in which the \hi\ alone does not effectively probe the highest gas densities where star formation is occurring.}  
\begin{figure*}
	\includegraphics[width=2.15\columnwidth]{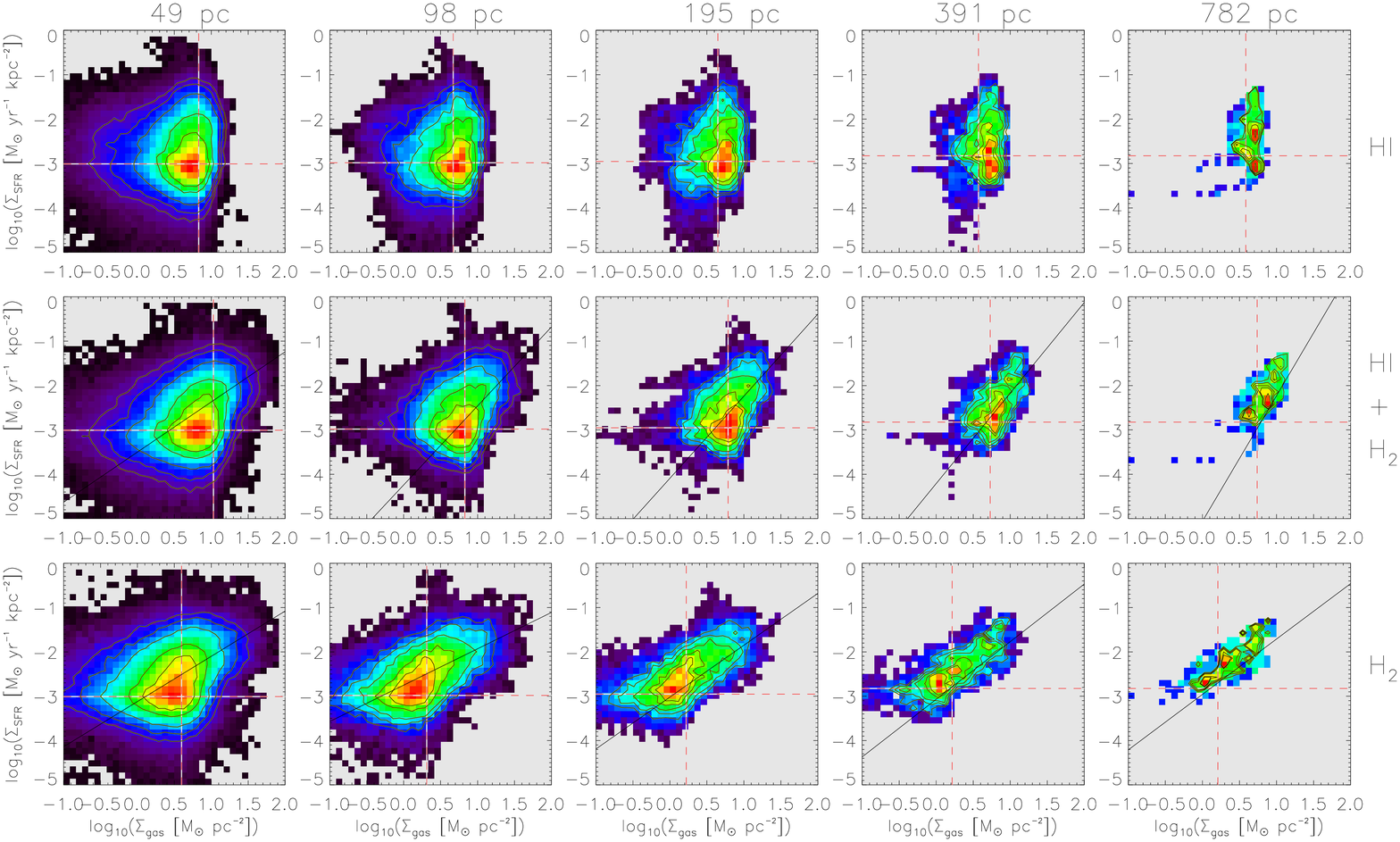}
	\caption{Total SFR surface density, \SFRSDhybrid, shown as functions of neutral gas surface density, $\Sigma_\mathrm{gas}$.  From top to bottom, the rows represent the cases in which $\Sigma_\mathrm{gas}=\Sigma_\mathrm{HI},~\Sigma_\mathrm{HI+H_2},~\Sigma_\mathrm{H_2}$.  From left to right, the columns represent the surface density maps at spatial resolutions of 49, 98, 195, 391, 782~pc.  In each panel, the horizontal and vertical red-dashed lines represent, respectively, the 3$\sigma$ surface densities of the corresponding $\Sigma_\mathrm{gas}$ and $\Sigma_\mathrm{SFR_{W3}}$ maps.  The reliable portion of each 2-dimensional distribution is therefore the upper right quadrant.  The black contours in each panel mark the 60, 70, 80, 90, and 95~percent percentiles of the number of points per two-dimensional bin.  The solid black lines shown in the panels in the middle and bottom rows do \emph{not} represent any sort of fit to the data.  Rather, they are the Kennicutt-Schmidt star formation laws fit to the red-filled circles shown in the corresponding panels in Fig.~\ref{fig:SFRW3_laws}.}
    \label{fig:SFRtotal_laws}
\end{figure*}

For completeness, we have also carried out pixel-pixel comparisons of $\log_{10}\Sigma_\mathrm{gas}$ and $\log_{10}\Sigma_\mathrm{SFR_{FUV+24~\mu m}}$.  The resulting star formation laws are presented in Fig.~\ref{fig:SFRtotal_laws} in the same manner as those shown in Fig.~\ref{fig:SFRW3_laws}.  Linear relationships between the two quantities exist when considering the \hi\ +\htwo\  and \htwo\ components of the neutral gas, yet they are not as well-defined as those based on the 12~\micron\ monochromatic SFR estimator.  At smaller length scales (e.g., 49 and 98~pc) the linear relationships seen at larger length scales break down considerably, yet are still present (albeit with much more scatter).  Thus, in addition to the 12~\micron\ monochromatic SFR estimator serving as an accurate tracer of the total SFR, it also yields better-defined Kennicutt-Schmidt star formation laws

\subsection{Discussion}
\citet{Onodera_2010}  studied the relationship between the surface densities of molecular gas and SFR in M~33 down to a spatial resolution of $\sim~80$~pc.  They used a combination of H$\alpha$ and 24~\micron\ imaging to estimate the SFR and found it to correlate well with \htwo\ surface densities at scales of  $~\sim~1$~kpc.  However, they show the correlation to break down when approaching giant molecular cloud (GMC) length scales.  \citet{Onodera_2010} attribute the break down to the variety of star formation activity among GMCs, which they in turn attribute to the various evolutionary stages of GMCs and to the drift of young star clusters from their parent GMCs.  They conclude that the Kennicutt-Schmidt star formation law is valid only on length scales greater than those of the parent GMCs.  Results similar to those of \citet{Onodera_2010} were found by \citet{Schruba_2010} who also studied the star formation law in M~33 over a range of scales.  On large (kpc) scales, they found a molecular star formation law described by a power law with index in the range $\sim$~1.1 to 1.5.  However, when moving to smaller scales, \citet{Schruba_2010} report a break down of the star formation law.  \citet{Schruba_2010} did, however, focus their analysis only on the regions in M~33 with the highest \Halpha\ and CO fluxes.

In this work, using WISE 12~\micron\ emission as a monochromatic SFR tracer, we find no evidence for a significant breakdown in the Kennicutt-Schmidt star formation law (especially the \htwo\ version) at GMC length scales in M~33.  Our results may be most directly compared to those of \citet{Williams_2018} who have also studied the star formation law in M~33 on sub-kpc length scales.  \citet{Williams_2018} used a variety of star formation tracers (including FUV~+~24~\micron) to determine the star formation rate in M~33.  To trace the gas content, they used the same \hi\ and CO image sets we use in this work.  While they do find clear power law correlations between $\Sigma_\mathrm{SFR}$ and $\Sigma_\mathrm{H_2}$, their indices are significantly higher than unity.  At spatial scales of 100, 400, and 1000~pc, they find $N=2.21\pm 0.05,~1.69\pm0.09,~\mathrm{and}~1.30\pm 0.11$, respectively.  It is only at length scales of $\sim 2$~kpc that they begin to recover a linear correlation between $\Sigma_\mathrm{SFR}$ and $\Sigma_\mathrm{H_2}$.  Like us, they also find a scale dependence of the power law index.  This, they say, indicates that the GMCs in M~33 are in a variety of evolutionary states.  However, for the $\Sigma_\mathrm{gas}=\Sigma_\mathrm{H_2}$ case, while our power law indices increase with spatial scale, theirs decrease.  Interestingly, their power law indices for the total gas cases generally increase at larger length scales, as do ours. 


\subsection{Star formation efficiencies}
In this section, we combine our SFR and gas surface density maps to produce star formation efficiency maps, $\mathrm{SFE}=\Sigma_\mathrm{SFR}/\Sigma_\mathrm{gas}$.  Having units of yr$^{-1}$, the SFE maps provide a measure of the time required for the current level of star formation to consume the observed gas supply (i.e., the gas depletion time).  For each different spatial scale, we generate SFE maps based on the \hi\ gas (SFE$_\mathrm{HI}=\Sigma_\mathrm{SFR}/\Sigma_\mathrm{HI}$), the total gas (SFE$_\mathrm{HI+H_2}=\Sigma_\mathrm{SFR}/\Sigma_\mathrm{HI+H_2}$), and the \htwo\ gas (SFE$_\mathrm{H_2}=\Sigma_\mathrm{SFR}/\Sigma_\mathrm{H_2}$).  Our first set of SFE maps are based on our 12~$\mu$m-based SFR maps, these  are shown in Fig.~\ref{SFEW3_vs_gas}.  For completeness, we also use our FUV+24~$\mu$m SFR maps to produce a set of SFE maps, these are shown in Fig.~\ref{SFEtotal_vs_gas}.

\begin{figure*}
	\includegraphics[width=2.15\columnwidth]{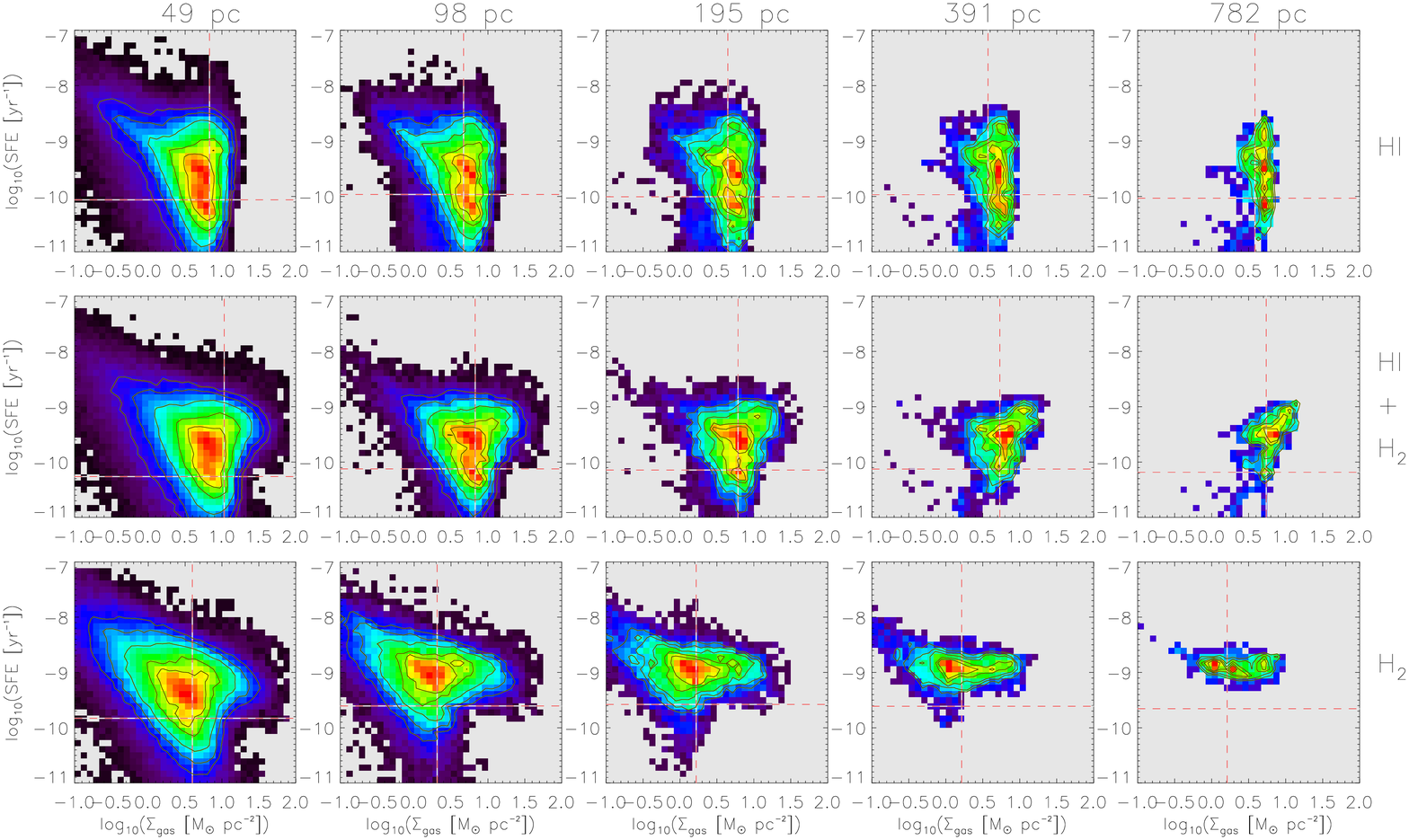}
	\caption{12~\micron-derived star formation efficiency maps shown as functions of neutral gas surface density, $\Sigma_\mathrm{gas}$.  From top to bottom, the rows represent the cases in which $\Sigma_\mathrm{gas}=\Sigma_\mathrm{HI},~\Sigma_\mathrm{HI+H_2},~\Sigma_\mathrm{H_2}$.  From left to right, the columns represent the surface density maps at spatial resolutions of 49, 98, 195, 391, 782~pc.  In each panel, the horizontal and vertical red-dashed lines represent, respectively, the 3$\sigma$ surface densities of the corresponding $\Sigma_\mathrm{gas}$ and SFE maps.  The reliable portion of each 2-dimensional distribution is therefore the upper right quadrant.  The black contours in each panel mark the 60, 70, 80, 90, and 95~percent percentiles of the number of points per two-dimensional bin.  \textcolor{black}{The bottom panels clearly show that at each of the length scales considered in this study, the \htwo-based SFEs have values close to $\sim 10^{-9}$~yr$^{-1}$ for all \htwo\ mass surface densities above  3$\sigma$ - they do not vary significantly with length scale.}}
    \label{SFEW3_vs_gas}
\end{figure*}

\begin{figure*}
	\includegraphics[width=2.15\columnwidth]{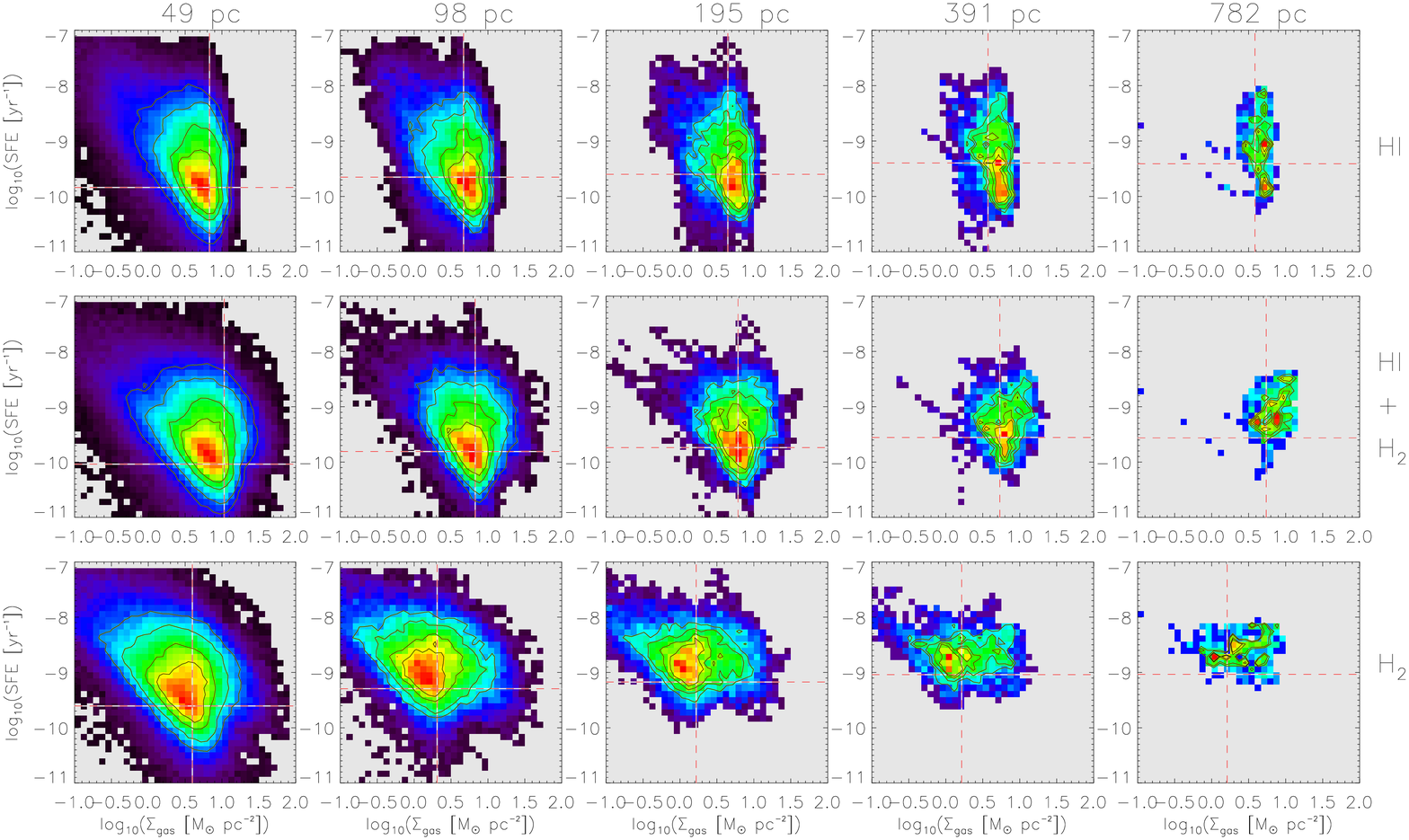}
	\caption{FUV+24~\micron-derived star formation efficiency maps shown as functions of neutral gas surface density, $\Sigma_\mathrm{gas}$.  From top to bottom, the rows represent the cases in which $\Sigma_\mathrm{gas}=\Sigma_\mathrm{HI},~\Sigma_\mathrm{HI+H_2},~\Sigma_\mathrm{H_2}$.  From left to right, the columns represent the surface density maps at spatial resolutions of 49, 98, 195, 391, 782~pc.  In each panel, the horizontal and vertical red-dashed lines represent, respectively, the 3$\sigma$ surface densities of the corresponding $\Sigma_\mathrm{gas}$ and SFE maps.  The reliable portion of each 2-dimensional distribution is therefore the upper right quadrant.  The black contours in each panel mark the 60, 70, 80, 90, and 95~percent percentiles of the number of points per two-dimensional bin.}
    \label{SFEtotal_vs_gas}
\end{figure*}

The ways in which star formation efficiency varies with gas mass on molecular cloud length scales are not well understood (\citealt{Kennicutt_Evans_2012}, and references therein).  Our SFE maps, especially those based on the 12~\micron\ SFRs, can be used to try to \textcolor{black}{address} this question for the case of M~33.  The bottom panels of Fig.~\ref{SFEW3_vs_gas} clearly show that at each of the length scales considered in this study, the \htwo-based SFEs have values close to $\sim 10^{-9}$~yr$^{-1}$ for all \htwo\ mass surface densities above  3$\sigma$.  \textcolor{black}{The spread in SFE$_\mathrm{H_2}$ values certainly does increase as the length scale decreases: from right to left in the bottom row of Fig.~\ref{SFEW3_vs_gas}, the standard deviations of the Gaussian-distributed SFE$_\mathrm{H_2}$ values that have corresponding $\Sigma_\mathrm{SFR_{W3}}$ values above the 3$\sigma$ thresholds are 0.15, 0.19, 0.28, 0.36, 0.52 dex, respectively}.   This could be due to the manner in which a larger range of GMC evolutionary states is resolved by the smaller length scales, whereas the range of states is spatially averaged over the larger length scales.  However, much like the correlations between $\Sigma_\mathrm{SFR}$ and $\Sigma_\mathrm{gas}$ shown in the previous section, the dominant mean value of SFE$_\mathrm{H_2}\sim 10^{-9}$~yr$^{-1}$ seen at lower spatial resolutions definitely does persist to the higher resolutions.  Therefore, for a large portion of the star-forming disc of M~33, the \htwo-based SFEs do seem to remain fairly constant over a range of length scales that includes molecular cloud length scales ($\sim 20$~-~$100$~pc in effective radius, \citealt{Gratier_2012}).

The relatively tight, horizontal distributions of SFEs seen in the \htwo\ cases disappear when considering the total gas surface densities and the \hi\ surface densities.  In fact, the situation is largely reversed - a large range of SFEs correspond to a relatively small range of gas surface densities.  Hence, the manners in which SFEs vary with gas mass clearly depend very much on the particular gas tracer being utilised.  They also depend on the SFR estimator.  In the bottom panels of Fig.~\ref{SFEtotal_vs_gas}, the FUV+24~\micron-based SFEs generally span a larger range of values than they do for the 12~\micron-based case.

\citet{Williams_2018} also investigated the length scale dependence of the star formation efficiency in M~33.  They find no significant variation in SFE with scale, for all three of their gas tracers.  Our results and those of \citet{Williams_2018} are in contradiction with those of \citet{Schruba_2010} who found a variation in gas depletion time with length scale in M~33.  However, as clearly discussed by \citet{Schruba_2010}, their results (including their measures of the star formation law at various length scales) may be affected by the manner in which their study targeted the regions in M~33 with the highest \Halpha\ and CO fluxes, rather than the full range of reliable fluxes.

\section{Star formation thresholds}\label{SF_thresholds}
The maps presented in Fig.~\ref{fig:maps} can  be used to test various star formation thresholds that attempt to link the observed star formation in a galaxy to the presence of gravitationally bound, cold gas clouds capable of collapsing into stars.  In this section, we test two of the more common star formation thresholds.  We do this only for the full-resolution (i.e., 49~pc) maps of the galaxy.  We do not study the thresholds as a functions of spatial scale. 

\subsection{Toomre model}\label{toomre}
\citet{Toomre_1964} showed that a stellar disc that is fairly smooth or uniform, and that is rotating in approximate equilibrium between its self-gravitational and centrifugal forces, cannot be entirely stable against the tendency to gravitationally collapse.  The Toomre criterion is most commonly used to quantify the gravitational growth of perturbations within a thin, rotating gaseous disc.  It describes the ability of perturbations to rotate about their centre of gravity and thus their stability against gravitational collapse.  According to this criterion, the disc should be unstable to axisymmetric disturbances in regions where the Toomre parameter, 
\begin{equation}
Q_\mathrm{gas}= {\alpha_\mathrm{Q}\sigma_\mathrm{gas}\kappa\over \pi G\Sigma_\mathrm{gas}},
\label{Toomre}
\end{equation}
is less than unity.  The self-gravity, pressure and kinematics of the gas disc are represented by $\Sigma_\mathrm{gas}$, $\sigma_\mathrm{gas}$ (velocity dispersion), and $\kappa$ (epi-cyclic frequency), respectively.  $G$ is Newton's gravitational constant.  $\alpha_\mathrm{Q}=0.69$ is an empirical calibration factor introduced by \citet{Martin_Kennicutt_2001} who used a sample of 32 star-forming spirals galaxies to show that the edges of their discs correspond to a median value of $Q_\mathrm{gas}=1.5$, not unity.  

\subsubsection{Generating the maps}
Our aim in this section is to create a \qgas\ map for M~33 using the full amounts of information contained in the full-resolution \hi\ data cube.  To this end, in addition to the $\Sigma_\mathrm{HI}$ map shown in Fig.~\ref{fig:maps}, we also require 2-dimensional velocity dispersion and epi-cyclic frequency maps.  In order to generate an \hi\ velocity dispersion map, we calculated the second-order moments of the third-order Gauss-Hermite polynomials fit to the line profiles in the \hi\ data cube (cf.~Section~\ref{Section_Data}).  This map is shown in the left panel of Fig.~\ref{HI_VF_dispersion}.  As will be shown below, the our 2-dimensional epi-cyclic frequency map, $\kappa(x,y)$, is produced using the \hi\ velocity field of the galaxy together with the parameters from a model fit to the \hi\ rotation curve.  In order to generate an \hi\ velocity field, we used the central velocities of the third-order Gauss-Hermite polynomials fit to the line profiles in the \hi\ data cube.  Our \hi\ velocity field for M~33 is shown in the right panel of Fig.~\ref{HI_VF_dispersion}.

\begin{figure*}
	\includegraphics[width=1.8\columnwidth]{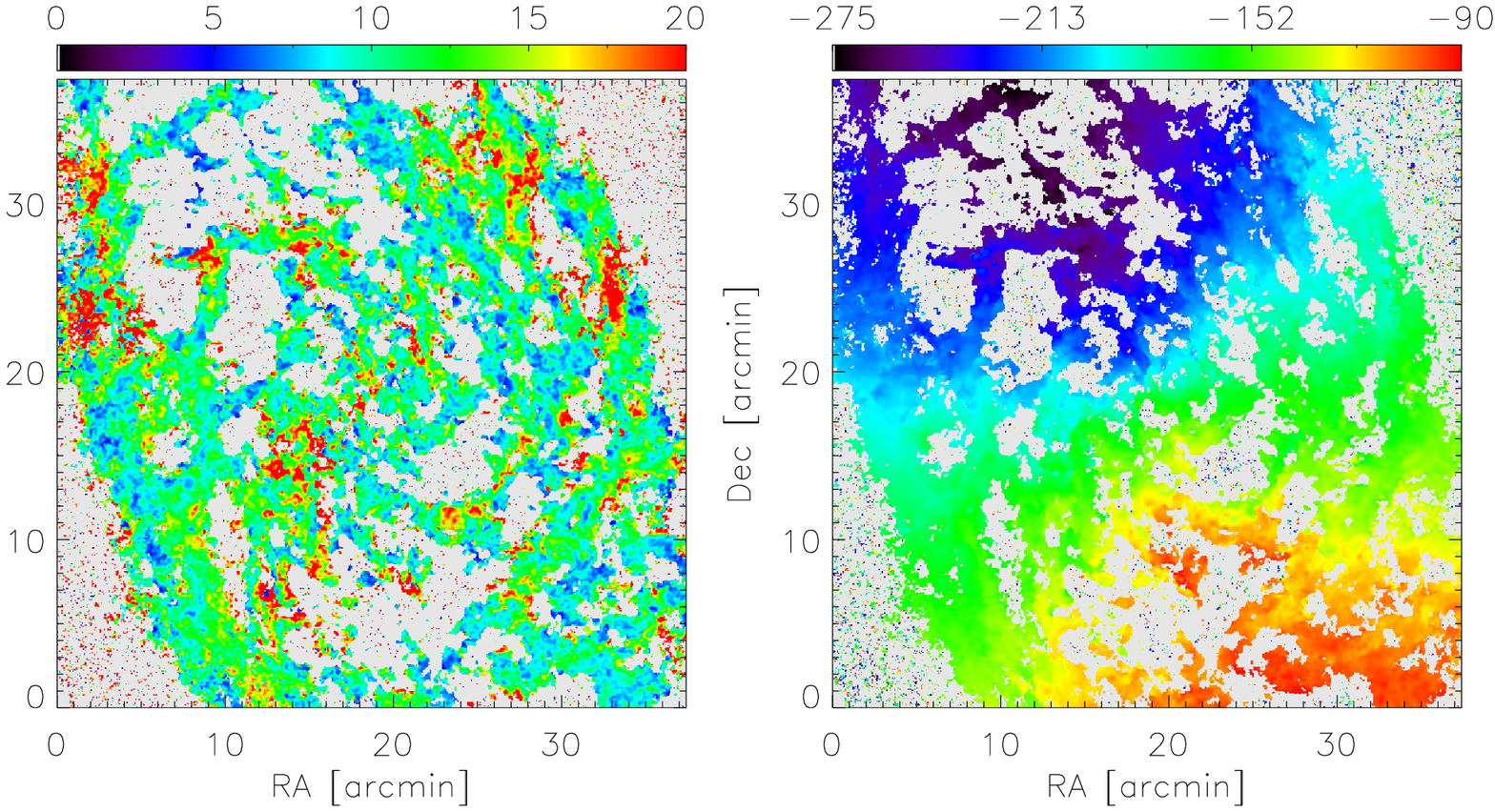}
    \caption{\hi\ velocity dispersion map (left) and \hi\ velocity field (right) for the full-resolution (49~pc) \hi\ data cube.  Pixels with corresponding values less than the 1$\sigma$ noise level (4~\msunppc) in the \hi\ total intensity map ($\Sigma_\mathrm{HI}$) have been blanked.  The colour bar above each panel specifies the velocities in units of \kms.  These maps, together with the $\Sigma_\mathrm{HI}$ map and the epi-cyclic frequency  map, $\kappa(x,y)$, are used to generate the instability maps shown in Fig.~\ref{fig:toomre1} and Fig.~\ref{fig:shear}.}
    \label{HI_VF_dispersion}
\end{figure*}

The Coriolis or centrifugal forces caused by the rotation of perturbations is approximated by the epi-cyclic frequency, $\kappa$.  Following \citet{Kennicutt_1989}, $\kappa$ is calculated as 
\begin{equation}
\kappa(R)=1.41{V(R)\over R}\sqrt{1+{R\over V(R)}{dV(R)\over dR}}, 
\label{kappa}
\end{equation}
where $V(R)$ is the observed rotational velocity.  In this work, in order to generate a 1-dimensional profile of the epi-cyclic frequency, $\kappa(R)$, we use the M~33 rotation curve presented in \citet{Kam_2017} based on \hi\ line observations obtained with the synthesis telescope at the Dominion Radio Astrophysical Observatory.  \citet{Kam_2017} fit a tilted ring model \citep{Rogstad_1974} to the \hi\ velocity field of M~33, they present the results in their Table~4.  Here, we fit their measured rotation curve with the analytic function
\begin{equation}
V_\mathrm{PE}(R)=V_0\left(1-e^{-R/R_\mathrm{PE}}\right)\left(1+{\alpha R\over R_\mathrm{PE}}\right).
\label{polyex_eqn}
\end{equation}
This is the Polyex model from \citet{Giovanelli_2002}.  $V_0$, $R_\mathrm{PE}$, and $\alpha$ determine the amplitude of the outer rotation curve, the exponential scale length of the inner rotation curve, and the slope of the outer rotation curve.  Our best-fit parameters are $V_0~=~96.8$~\kms, $R_\mathrm{PE}~=~4.9$~arcmin, and $\alpha~=~0.018$.  Figure~\ref{fig:Vrot} (left panel) shows the \citet{Kam_2017} measured rotation curve together with the best-fit Polyex model.  

\begin{figure*}
	\includegraphics[width=2\columnwidth]{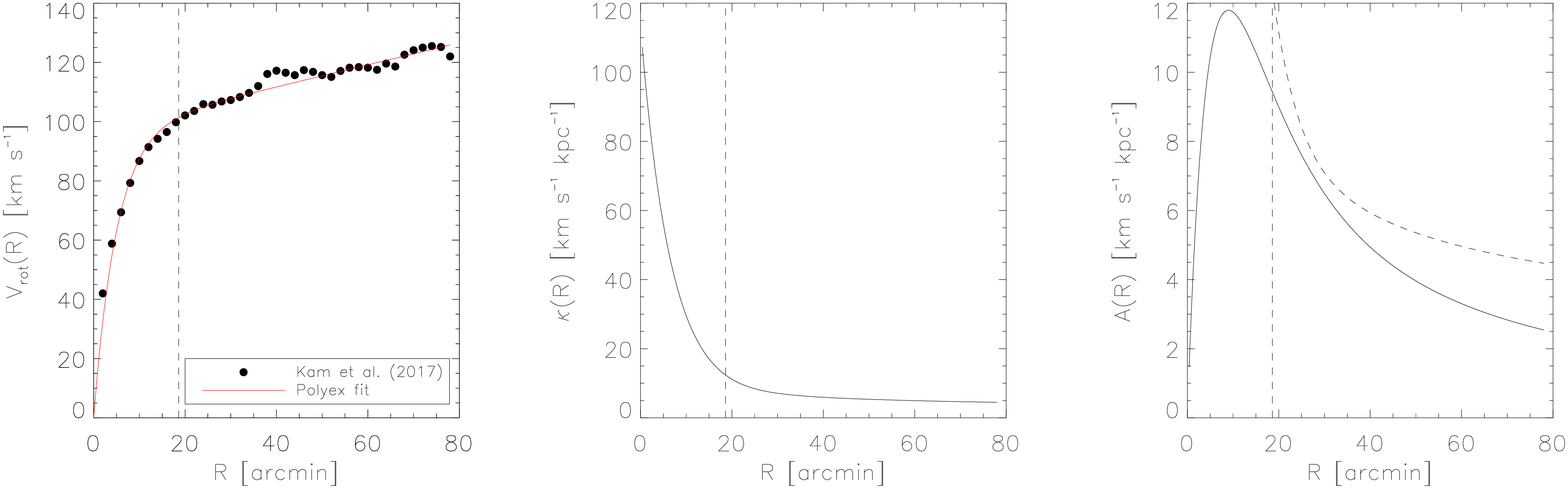}
    \caption{Left: Rotation curve for M~33 from \citet{Kam_2017} (black-filled circles) parameterised using the Polyex model (red curve).  Middle: Epi-cyclic frequency radial profile, $\kappa(R)$, for the Toomre \qgas\ model based on best-fit parameters from the Polyex rotation curve.  Right: Shear profile for the shear-based model of \citet{Hunter_1998} (solid curve) based on best-fit parameters from the Polyex rotation curve.  For comparison, the epi-cyclic frequency profile from the middle panel is shown as the dashed curve.  \textcolor{black}{Clearly, $\kappa(R)$ and $A(R)$ are significantly different to one another over the steeply rising portion of the rotation curve.  At outer radii, they approach similar values.}   The vertical dashed line in each panel is at a radius $R=18.6$~arcmin, it represents the maximum radius out to which the star formation thresholds are studied.  This outer radius is set by the physical extent of the CO disc as seen in the $^{12}$CO($J$~=~2-1) total intensity map from \citet{Druard_2014}.}
    \label{fig:Vrot}
\end{figure*}

Using the Polyex model for the rotation curve, Eqn.~\ref{kappa} can be re-written as
\begin{equation} \label{kappa2}
\begin{split}
\kappa(R)=1.41{V_\mathrm{PE}(R)\over R}&\Bigg(1+{V_oR\over V_\mathrm{PE}(R)}e^{-R/R_\mathrm{PE}}\left({1\over R_\mathrm{PE}}+{\alpha R\over R_\mathrm{PE}^2}+{1\over R}\right)\\
& - {V_0\over V_\mathrm{PE}(R)}\Bigg)^{1/2}.
\end{split}
\end{equation}
$\kappa(R)$ based on the best-fit Polyex model parameters is shown in the middle panel of Fig.~\ref{fig:Vrot}.

In order to convert the 1-dimensional $\kappa(R)$ profile shown in Eqn.~\ref{kappa2} into a 2-dimensional map, $\kappa(x,y)$, that incorporates the information contained in the \hi\ velocity field of M~33, we replace $V_\mathrm{PE}(R)$ in Eqn.~\ref{kappa2} with 
\begin{equation}
V_\mathrm{PE}(x,y)={VF(x,y)-V_\mathrm{sys}\over \sin i \cos\theta},
\end{equation} 
where $VF(x,y)$ is the radial velocity at position $(x,y)$ as given by the velocity field, $V_\mathrm{sys}=-180$~\kms \citep{Kam_2017}, $i=53.9$\deg, and $\theta=201.3$\deg.  The tilted ring model from \citet{Kam_2017} shows the position angle of the galaxy to be constant at a value of 201.3\deg\ out to radius $R=30$~arcmin.  Over the same radial range, the inclination varies linearly from a value of 53.2\deg\ at the centre to 54.6\deg.  When calculating the various star formation thresholds in this section, we are restricted to the portion of the galaxy within a radius $R\approx 18.6$~arcmin.  This radial range very much constitutes the rising portion of the rotation curve.   This limit is imposed but the physical extent of the CO image from \citet{Druard_2014} used to generate the \HIISD\ map.  This limiting radius is indicated by the black ellipses in Fig.~\ref{fig:maps}.  

Because the circular velocity components do not contribute to the line-of-sight velocities along the minor axis of the \hi\ velocity field ($\cos\theta=0$), all pixels in the \qgas\ maps within 10\deg\ of the minor axis are ignored (i.e., blanked).  Furthermore, pixels in the relevant $\Sigma_\mathrm{gas}$ map that are below the 3$\sigma$ level are also blanked.  Hence, the three \qgas\ maps (and the \sgas\ maps in the next section) all have different filling factors.  For the \qgas\ map based on $\Sigma_\mathrm{gas}=\Sigma_\mathrm{H_2}$, only pixels corresponding to $\Sigma_\mathrm{H_2}>10.2$~\msun~pc$^{-2}$ are considered.  Similarly, only pixels with $\Sigma_\mathrm{H_2}>10.2$~\msun~pc$^{-2}$ and $\Sigma_\mathrm{HI}>4$~\msun~pc$^{-2}$ are used for the \qgas\ map based on  $\Sigma_\mathrm{gas}=\Sigma_\mathrm{HI+H_2}$.  The \qgas\ map based on  $\Sigma_\mathrm{gas}=\Sigma_\mathrm{HI}$ uses only those pixels with  $\Sigma_\mathrm{HI}>4$~\msun~pc$^{-2}$.  For the instability maps based only on the H$_2$ component of the gas, we scale the \hi\ velocity dispersion map by a factor of 1/1.4.  This is based on the results from \citet{Mogotsi_2016} who show the ratio of \hi\ to CO velocity dispersions in THINGS galaxies to be 1.4.

\subsubsection{Results and Discussion}
Setting $Q_\mathrm{gas}=1$ in Eqn.~\ref{Toomre} and solving for $\Sigma_\mathrm{gas}$ yields a \textit{critical} gas surface density, $\Sigma_\mathrm{crit}$, above which the gas should be gravitationally unstable.  Figure~\ref{fig:toomre1} shows maps of $\Sigma_\mathrm{gas}/\Sigma_\mathrm{crit}$ for the cases $\Sigma_\mathrm{gas}~=~\Sigma_\mathrm{HI}$, $\Sigma_\mathrm{gas}~=~\Sigma_\mathrm{HI+H_2}$, and $\Sigma_\mathrm{gas}~=~\Sigma_\mathrm{H_2}$.  For all cases, the Toomre model significantly under-predicts the amount of neutral gas over large inner portions of the galaxy.  For the \hi-only case, no star formation is predicted to occur within a radius $R\sim 8.5$~arcmin of the centre.  The results are similar for the \hi+\htwo\ case.  However, including the \htwo\ yields higher gas surface densities that result in only the inner-most disc ($R\lesssim 3$~arcmin) being incorrectly predicted to have gas surface densities too low for star formation.  The persistent inability of the \qgas\ to correctly predict the star formation near the galaxy centre is presumably due to the very high epi-cyclic frequencies occurring at small radii, associated with the steeply rising portion of the rotation curve.  At $R\sim 3$~arcmin, $\kappa\sim 80$~\kms~kpc$^{-1}$ (see middle panel of Fig.~\ref{fig:Vrot}).  For the \htwo-only case, the Toomre model again incorrectly predicts the inner few arcmin of the galaxy to have no star formation.  However, unlike the other two cases, it does a  good job of predicting fairly well the observed \htwo\ surface densities over the rest of the disc ($R\gtrsim 3.5$~arcmin).  Over this entire radial range, the mean value of $\Sigma_\mathrm{H_2}/ \Sigma_\mathrm{crit}$  varies slowly from $\sim 1$ to $\sim 4$.  

Overall, the Toomre model consistently predicts a lack of star formation in the inner-most parts of M~33.  This result may be compared to the findings of \citet{THINGS_Leroy} who for samples of spirals and dwarfs from THINGS found \qgas\ to suggest the galaxies to always be stable against the tendency to form stars (regardless of galacto-centric radius).  Similar conclusions were reached by \citet{Elson_2012} for the nearby blue compact dwarf galaxies NGC~2915 and NGC~1705.  Hence, for the case of M~33, \qgas\ does a relatively better job of at least predicting the outer portions of the star-forming disc to be gravitationally unstable. 

\begin{figure*}
	\includegraphics[width=2\columnwidth]{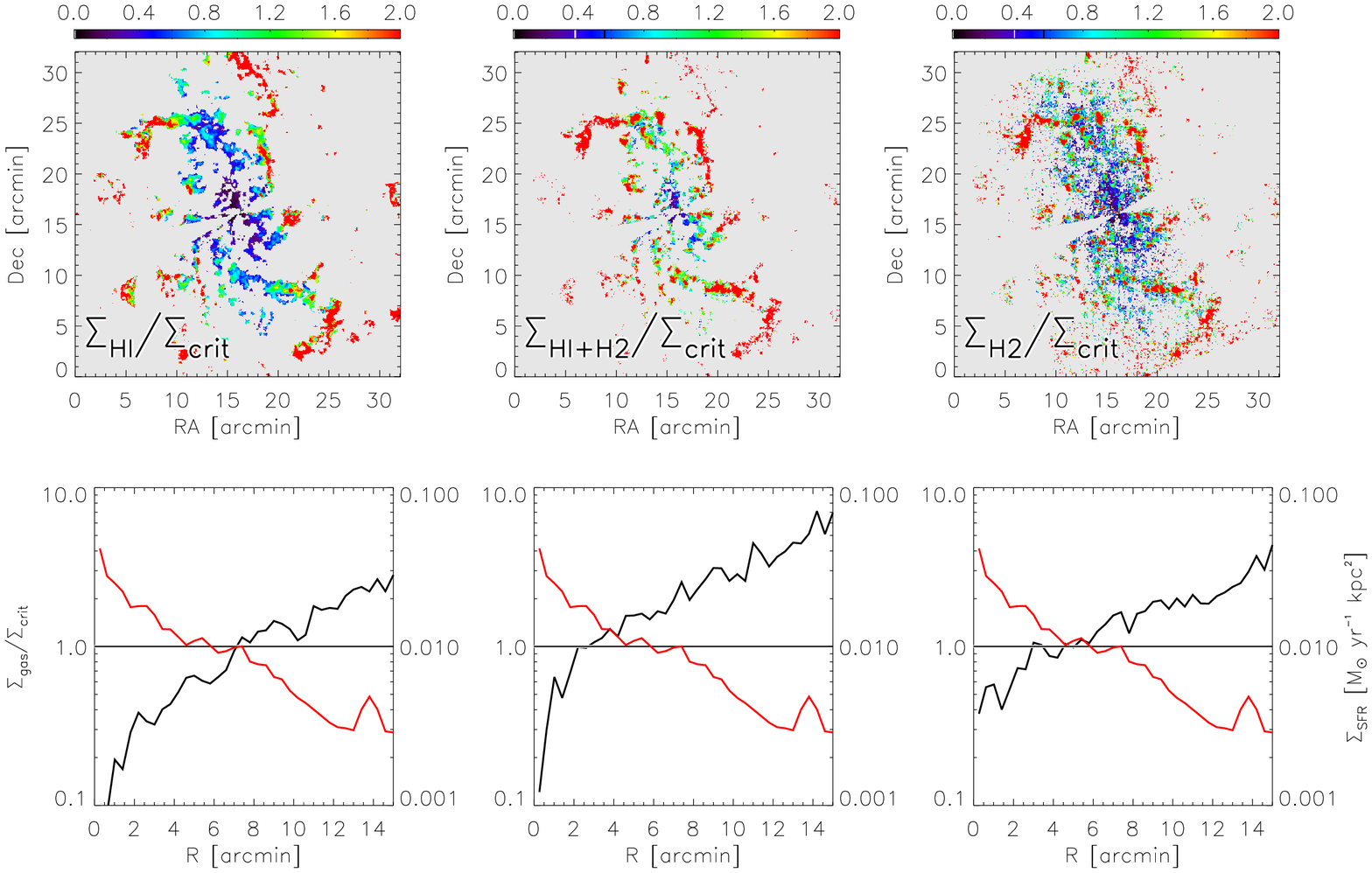}
    \caption{\qgas\ star formation threshold results for M~33 for the cases $\Sigma_\mathrm{gas}=\Sigma_\mathrm{HI}$ (left), $\Sigma_\mathrm{gas}=\Sigma_\mathrm{HI+H_2}$ (middle), and $\Sigma_\mathrm{gas}=\Sigma_\mathrm{H_2}$ (right).  The top row shows maps of the ratio of the observed gas density to the critical gas density required for star formation, $\Sigma_\mathrm{gas}/\Sigma_\mathrm{crit}$. Bottom panels show as solid curves the median values of $\Sigma_\mathrm{gas}/\Sigma_\mathrm{crit}$ (black) as well as the median SFR surface densities (red).  Each panel contains two different y-axes.  On the left is $\Sigma_\mathrm{gas}/\Sigma_\mathrm{crit}$ while on the right is \SFRSD.  In each panel, the horizontal black line marks $\Sigma_\mathrm{gas}/\Sigma_\mathrm{crit}=1$.  \textcolor{black}{For all cases, the Toomre model significantly under-predicts the amount of neutral gas over large inner portions of the galaxy.}}
    \label{fig:toomre1}
\end{figure*}

\subsection{Shear model}\label{hunter_shear}
\citet{Hunter_1998} test the efficacy of various star formation thresholds to predict the relationship between gas, stars and star formation in a sample of irregular galaxies.  They consider the situation in which cloud growth occurs with streaming motions along interstellar magnetic field lines.  Under such circumstances, they say, the Coriolis force can be less important than the time available for clouds to grow in the presence of gravitational shear.  They quantify the rotational shear of the gas using Oort's $A$ constant,
\begin{equation}
A(R)=0.5\left({V(R)\over R} - {dV(R)\over dR}\right).
\end{equation}
Then, their shear-based parameter for gravitational instability is 
\begin{equation}
S_\mathrm{gas}={\alpha_A\sigma_\mathrm{gas}A\over \pi G\Sigma_\mathrm{gas}}.  
\end{equation}

\citet{Hunter_1998} estimate $\alpha_A=2.5$.  This value matches the contrast between the surface densities of neutral and molecular inter-stellar media in the presence of rotational shear.  Regions in the galaxy with $S_\mathrm{gas}\le 1$ should be unstable against the tendency to gravitationally collapse (i.e., forming stars), while regions with $S_\mathrm{gas}>1$ should be stable against large-scale gravitational collapse.  The right panel of Fig.~\ref{fig:Vrot} shows the radial profile of $A$ based on the Polyex parameterisation of the M~33 rotation curve. Clearly,  $A << \kappa$ at inner radii  while at outer radii ($R\gtrsim 20$~arcmin) $A \approx\kappa$.  \citet{Hunter_1998} stress that using $A$ instead of $\kappa$ to quantify the kinematics  makes very little difference for the flat portion of a rotation curve.  For the rising portion of a rotation curve where the shear is low, $A<<\kappa$ results in $S_\mathrm{gas}<<Q_\mathrm{gas}$.  

\subsubsection{Generating the maps}
Using the functional form of the Polyex model given in Eqn.~\ref{polyex_eqn} to calculate a 1-dimensional profile of the rotational shear yields
\begin{equation}
A(R)=0.5\left( {V_0\over R} -V_0 e^{-r/R_\mathrm{PE}}\left( {1\over R_\mathrm{PE}} + {\alpha R\over R_\mathrm{PE}^2} + {1\over R}  \right) \right).
\end{equation}
We were unable to express $A(R)$ in terms of $V_\mathrm{PE}(R)$.   Given that $A(R)$ depends only on the Polyex model parameters, our shear-based instability maps are not based on the rotational velocities sourced directly from the \hi\ velocity field of the galaxy.  Instead, they are based on $V(R)$ (via the Polyex model parameterisation).  Our 2-dimensional $A(x,y)$ map is therefore a radially symmetric map.    It is, however, combined with the non-symmetric maps of the \hi\ velocity field and \hi\ velocity dispersion, thereby yielding $S_\mathrm{gas}$ maps that do indeed vary with (x,y) position.

\subsubsection{Results and discussion}
Figure~\ref{fig:shear} shows maps of $\Sigma_\mathrm{gas}/\Sigma_\mathrm{crit}$ for the cases $\Sigma_\mathrm{gas}~=~\Sigma_\mathrm{HI}$, $\Sigma_\mathrm{gas}~=~\Sigma_\mathrm{HI+H_2}$, and $\Sigma_\mathrm{gas}~=~\Sigma_\mathrm{H_2}$.  Unlike the \qgas\ models, all three \sgas\ threshold maps predict the presence of star formation in the innermost regions of the M~33.  The \hi-only case formally predicts star formation only within $R\sim 2$~arcmin of the galaxy centre.  However, beyond this radius $\Sigma_\mathrm{HI}/\Sigma_\mathrm{crit}$ obtains at a roughly constant value of $\sim 0.5$.  For the total gas case, the shear model predicts the entire inner disc to be star-forming.  For $R\lesssim 6$~arcmin, the rate at which $\Sigma_\mathrm{gas}/\Sigma_\mathrm{crit}$ decreases with radius is very similar to the rate at which \SFRSDW3\ decreases.  This suggests the total gas mass plays an important role in regulating the gravitational instability of the interstellar medium.  The outer parts of the gas disc equilibrate to a roughly constant value of $\Sigma_\mathrm{HI+H_2}/\Sigma_\mathrm{crit}$, this time closer to unity than for the \hi-only case.  The shear model does not perform well for the \htwo-only case.  Like the \hi-only case, it predicts the innermost ($R\lesssim4$)~arcmin portion of the galaxy to be star-forming, and fails at outer radii.  Again, however,  $\Sigma_\mathrm{gas}/\Sigma_\mathrm{crit}$ equilibrates to a roughly constant value at larger radii.  

An instability parameter such as \sgas\ that equilibrates to a roughly constant value over an extended portion of the disc offers the exciting possibility of it being used to generate an estimate of the galaxy's rotation curve.  If a constant stability star-forming disc is assumed (e.g., \citealt{Meurer_2013}) and if the self-similarity of \hi\ column density profiles of galaxies (e.g., \citealt{Martinsson_2011}) is considered, the shape of a galaxy's rotation curve required to generate the constant instability parameter can be constrained.  Such a method of inferring the shapes of rotation curves will be well-suited to forthcoming \hi\ galaxy surveys to be carried out on ASKAP and MeerKAT that will directly detect many thousands of galaxy in \hi\ line emission, yet which will lack the resolution to spatially resolve them - thereby preventing the generation of rotation curves using traditional methods.

\begin{figure*}
	\includegraphics[width=2\columnwidth]{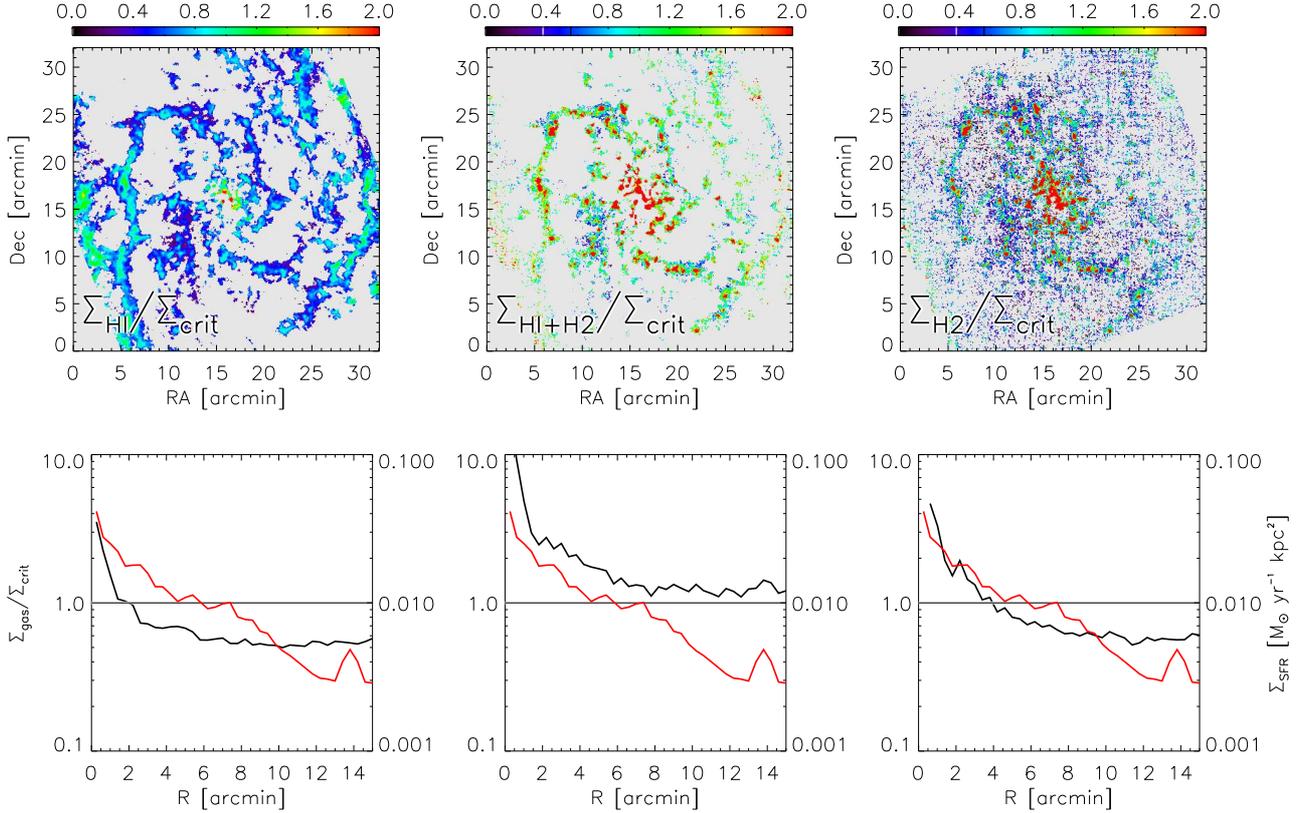}
    \caption{\sgas\ star formation threshold results for M~33.   See Fig.~\ref{fig:toomre1} for full details.  Note  for the $\Sigma_\mathrm{gas} = \Sigma_\mathrm{HI+H_2}$ and $\Sigma_\mathrm{gas} = \Sigma_\mathrm{H_2}$ cases that the rate at which $\Sigma_\mathrm{gas}/\Sigma_\mathrm{crit}$ decreases with radius is similar to the drop off in SFR.  \textcolor{black}{In all cases, the outer parts of the gas disc equilibrate to a roughly constant value of $\Sigma_\mathrm{gas}/\Sigma_\mathrm{crit}$}.}
    \label{fig:shear}
\end{figure*}

\section{\hi\ saturation limit}\label{saturation_limit}
A particular result from the study by \citet{THINGS_Bigiel} of  the star formation law in THINGS galaxies is the presense of a sharp saturation of \hi\ surface densities at $\approx 9$~\msunppc\ in both the spirals and dwarfs.  In the case of spirals, the gas in excess of this limit is observed to be molecular.  The THINGS data cubes have high spatial resolution ($\sim 6$~arcsec).  Given the distance range of the THINGS galaxies ($\sim 2$~-~15~Mpc), the corresponding physical resolutions are $\sim~50$~-~450~pc.  However, \citet{THINGS_Bigiel} degrade the physical resolution of all their \hi\ maps to 750~pc in order to carry out their study.  

The top panel of Fig.~\ref{fig:mom0_hist} shows the distribution of face-on \hi\ surface densities from the full-resolution \hi\ total intensity map of M~33 generated in this work (shown in Fig.~\ref{fig:maps}).  The median \hi\ mass surface density is 6.3~\msunppc.  Only $\sim 71$~percent of the \hi\ mass in M~33 is observed at surface densities  below the THINGS saturation limit of 9~\msunppc, the other 29~percent yields surface densities of up to 20~\msunppc, and higher.  At 12~arcsec angular resolution, the JVLA imaging has a corresponding physical resolution of $\approx 49$~pc.  This physical resolution is roughly similar to the intrinsic physical sizes of giant molecular clouds in M33, which \citet{Gratier_2012} showed to range from $\sim 20$~-~$100$~pc in effective radius.  Assuming the high-density \hi\ in the galaxy to be clumped on roughly similar length scales, it is reasonable to assume that the JVLA images are probing the true \hi\ surface densities of the galaxy.

\begin{figure}
	\includegraphics[width=\columnwidth]{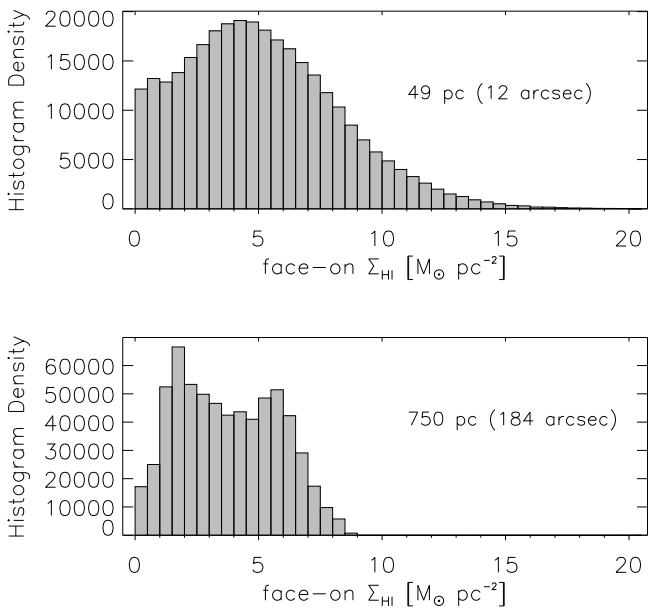}
    \caption{Distribution of \hi\ surface densities from the full-resolution (49~pc, 12~arcsec) \hi\ total intensity map of M~33 (top panel) and for the spatially-smoothed (750~pc, 184~arcsec) map (bottom panel).  All surface densities have been corrected for the inclination of the galaxy.  The same bin size of 0.5~\msunppc\ is used for both histograms.  \textcolor{black}{Spatially smoothing the data has the clear effect of artificially introducing an \hi\ saturation limit close to $\sim 9$~\msun~pc$^{-2}$.}}
    \label{fig:mom0_hist}
\end{figure}

The situation changes drastically when the JVLA \hi\ cube from \citet{Gratier_2010} is smoothed.  We used a Gaussian convolution kernel to smooth each channel of the \hi\ data cube to a spatial resolution of 750~pc.  The cube was then spectrally integrated to yield a total intensity map in units of \msunppc.  The bottom panel of Fig.~\ref{fig:mom0_hist} shows the distribution of the face-on \hi\ mass surface densities in the smoothed map.  For this smoothed version of the cube, the surface densities are distributed much more uniformly.  Most notable is that fact that the maximum \hi\ surface density has been reduced to 8.6~\msunppc,  very close to the saturation limit given by \citet{THINGS_Bigiel}.  

We carried out an identical experiment with one of the THINGS galaxies, NGC 1569.  The naturally-weighted \hi\ data cube for this galaxy has a spatial resolution of $7.71\times 7.04$~arcsec$^2$ (corresponding to a physical resolution of $\approx 74\times 68$~pc$^2$ at the distance of the galaxy, 2~Mpc).  In the \hi\ total intensity map, a significant portion of the disc has \hi\ mass surface densities well in excess of 9~\msunppc, some reaching as high as $\approx 32$~\msunppc.  When the cube is smoothed to a spatial resolution of 750~pc, the peak surface density in the \hi\ total intensity map drops to $\approx 9.7$~\msunppc.  

The simple demonstrations presented in this section suggests the sharp saturation of H\,{\sc i} surface densities at $\approx 9$~\msunppc\ reported by \citet{THINGS_Bigiel} for their 18 THINGS galaxies may be due to the manner in which the images were spatially smoothed. Galaxies such as M~33 and NGC~1569 observed at high spatial resolution have significant fractions of their \hi\ mass corresponding to surface densities higher than $\approx 9$~\msunppc.

\section{Summary}\label{summary}
In this work, we have used various existing image sets of M~33 to carry out several studies over a range of physical spatial resolutions spanning 49 to 782~pc.

We have shown the monochromatic star formation rate (SFR) estimator based on 12~\micron\ emission from polycyclic aromatic hydrocarbons to be reliably and accurately correlated to the traditional FUV + 24~\micron\ hybrid tracer of the total SFR.  Over the full range of length scales considered, a  linear correlation between $\Sigma_\mathrm{SFR_{W3}}$ and $\Sigma_\mathrm{SFR_{FUV+24~\mu m}}$ clearly persists, yet the Gaussian-distributed scatter in the correlation does increase as the length-scale is decreased.  We have therefore confirmed at sub-kpc length scales the similar results found by other authors on global length scales.  Furthermore, we have generated a global SFR estimate from the PACS 100~\micron\ imaging ($0.44\pm 0.10$~\msun~yr$^{-1}$) and shown is to be consistent with our 12~\micron-based estimate ($0.34^{0.42}_{0.27}$~\msun~yr$^{-1}$), thereby showing how the  WISE 12~\micron\ and  Herschel 100~\micron\ bands both trace the same star formation activity in the interstellar medium.  This further demonstrates the robust and credible nature of the WISE 12~\micron\ imaging and the methods used to calculate the SFR. 

We have addressed the question of how spatial sampling and averaging affect the observed form of the star formation law in M~33, and we have done this in terms of \hi, \htwo, and total gas surface densities.  Contrary to previously published results for M~33 from \citet{Onodera_2010} and \citet{Schruba_2010}, we find clear evidence for the prevalence of a Kennicutt-Schmidt star formation law over our full range of considered sub-kpc length scales.  For the case in which only the \htwo\ component of the neutral gas is considered, all correlations are entirely consistent or close to being consistent with a linear relation.  This is true all the way down to giant molecular cloud length scales.   When considering the total gas content (\hi~+~\htwo\ surface densities) of the galaxy,  the indices of the power-law correlations are all higher (closer to a value of $\approx 2$) than the corresponding values for the \htwo-only case.  No clear correlations are found for the \hi-only cases.  A similar extension of the Kennicutt-Schmidt law down to sub-kpc length scales in M~33 was recently found by \citet{Williams_2018} using the same gas maps used in this work, yet they find power law indices significantly higher than ours.

We combined our SFR and gas surface density maps to study the distribution of star formation efficiencies (SFEs) in M~33.  Over the full range of length scales considered, the \htwo-based SFEs are all centred on a mean value of $\sim 10^{-9}$~yr$^{-1}$.  We have therefore provided clear evidence for the existence of a fairly constant SFE in M~33, down to the length scales of molecular clouds.

We have used our high-resolution (49~pc) maps to study the Toomre instability threshold and the shear-based instability model of \citet{Hunter_1998}.  Our study is limited to the rising portion of the rotation curve.  The Toomre models consistently predict the inner parts of M~33 to have no star formation.  The shear-based models incorporating the total gas surface densities correctly predict the  entire inner disc to be star-forming.  In all cases, the shear-based models equilibrate to a roughly constant value of observed to critical gas surface densities at larger radii.  

Finally, we have discussed the possibility that observations of a saturation of \hi\ surface densities at $\sim 9$~\msunppc\ in nearby galaxies reported by other authors may be an artefact of the spatial smoothing process.  The 12~arcsec resolution \hi\ map used in this work  has only $\sim 71$~percent of its \hi\ mass surface densities below the THINGS saturation limit of $\sim 9$~\msunppc.  However, when the map is smoothed from its native resolution of 49~pc down to 750~pc (the same resolution as the THINGS imaging in \citealt{THINGS_Bigiel}), it does indeed clearly exhibit a maximum surface density close to $9$~\msunppc.  We carried out an identical test using one of the THINGS galaxies (NGC~1569), and found the same result.

\section{Acknowledgements}
ECE acknowledges that this research was supported by the South African Radio Astronomy Observatory, which is a facility of the National Research Foundation, an agency of the Department of Science and Technology.  This work is based on research supported in part by the National Research Foundation of South Africa (Grant Number 115238).  The work of LC is supported by the Comit\'e Mixto ESO-Chile and the DGI at University of Antofagasta.  The work of CC an TJ is based upon research supported by the South African Research Chairs Initiative (SARChI) of the Department of Science and Technology (DST),  South Africa SKA, and the National Research Foundation (NRF).  All authors sincerely thank the anonymous referee for providing insightful and constructive comments that surely improved the quality of this paper.


\bibliographystyle{mnras}

\begin{thebibliography}{}
\makeatletter
\relax
\def\mn@urlcharsother{\let\do\@makeother \do\$\do\&\do\#\do\^\do\_\do\%\do\~}
\def\mn@doi{\begingroup\mn@urlcharsother \@ifnextchar [ {\mn@doi@}
  {\mn@doi@[]}}
\def\mn@doi@[#1]#2{\def\@tempa{#1}\ifx\@tempa\@empty \href
  {http://dx.doi.org/#2} {doi:#2}\else \href {http://dx.doi.org/#2} {#1}\fi
  \endgroup}
\def\mn@eprint#1#2{\mn@eprint@#1:#2::\@nil}
\def\mn@eprint@arXiv#1{\href {http://arxiv.org/abs/#1} {{\tt arXiv:#1}}}
\def\mn@eprint@dblp#1{\href {http://dblp.uni-trier.de/rec/bibtex/#1.xml}
  {dblp:#1}}
\def\mn@eprint@#1:#2:#3:#4\@nil{\def\@tempa {#1}\def\@tempb {#2}\def\@tempc
  {#3}\ifx \@tempc \@empty \let \@tempc \@tempb \let \@tempb \@tempa \fi \ifx
  \@tempb \@empty \def\@tempb {arXiv}\fi \@ifundefined
  {mn@eprint@\@tempb}{\@tempb:\@tempc}{\expandafter \expandafter \csname
  mn@eprint@\@tempb\endcsname \expandafter{\@tempc}}}

\bibitem[\protect\citeauthoryear{{Bigiel}, {Leroy}, {Walter}, {Brinks}, {de
  Blok}, {Madore}  \& {Thornley}}{{Bigiel} et~al.}{2008}]{THINGS_Bigiel}
{Bigiel} F.,  {Leroy} A.,  {Walter} F.,  {Brinks} E.,  {de Blok} W.~J.~G.,
  {Madore} B.,   {Thornley} M.~D.,  2008, \mn@doi [\aj]
  {10.1088/0004-6256/136/6/2846}, \href
  {http://adsabs.harvard.edu/abs/2008AJ....136.2846B} {136, 2846}

\bibitem[\protect\citeauthoryear{{Boquien} et~al.,}{{Boquien}
  et~al.}{2010}]{Boquien_2010}
{Boquien} M.,  et~al., 2010, \mn@doi [\aap] {10.1051/0004-6361/201014649},
  \href {http://adsabs.harvard.edu/abs/2010A%26A...518L..70B} {518, L70}

\bibitem[\protect\citeauthoryear{{Brown} et~al.,}{{Brown}
  et~al.}{2017}]{Brown_2017}
{Brown} M.~J.~I.,  et~al., 2017, \mn@doi [\apj] {10.3847/1538-4357/aa8ad2},
  \href {http://adsabs.harvard.edu/abs/2017ApJ...847..136B} {847, 136}

\bibitem[\protect\citeauthoryear{{Calzetti} et~al.,}{{Calzetti}
  et~al.}{2005}]{Calzetti_2005}
{Calzetti} D.,  et~al., 2005, \mn@doi [\apj] {10.1086/466518}, \href
  {http://adsabs.harvard.edu/abs/2005ApJ...633..871C} {633, 871}

\bibitem[\protect\citeauthoryear{{Calzetti} et~al.,}{{Calzetti}
  et~al.}{2007}]{Calzetti_2007}
{Calzetti} D.,  et~al., 2007, \mn@doi [\apj] {10.1086/520082}, \href
  {http://adsabs.harvard.edu/abs/2007ApJ...666..870C} {666, 870}

\bibitem[\protect\citeauthoryear{{Cluver} et~al.,}{{Cluver}
  et~al.}{2014}]{Cluver_2014}
{Cluver} M.~E.,  et~al., 2014, \mn@doi [\apj] {10.1088/0004-637X/782/2/90},
  \href {http://adsabs.harvard.edu/abs/2014ApJ...782...90C} {782, 90}

\bibitem[\protect\citeauthoryear{{Cluver}, {Jarrett}, {Dale}, {Smith}, {August}
   \& {Brown}}{{Cluver} et~al.}{2017}]{Cluver_2017}
{Cluver} M.~E.,  {Jarrett} T.~H.,  {Dale} D.~A.,  {Smith} J.-D.~T.,  {August}
  T.,   {Brown} M.~J.~I.,  2017, \mn@doi [\apj] {10.3847/1538-4357/aa92c7},
  \href {http://adsabs.harvard.edu/abs/2017ApJ...850...68C} {850, 68}

\bibitem[\protect\citeauthoryear{{Druard} et~al.,}{{Druard}
  et~al.}{2014}]{Druard_2014}
{Druard} C.,  et~al., 2014, \mn@doi [\aap] {10.1051/0004-6361/201423682}, \href
  {http://adsabs.harvard.edu/abs/2014A%26A...567A.118D} {567, A118}

\bibitem[\protect\citeauthoryear{{Elmegreen}}{{Elmegreen}}{2018}]{Elmegreen_2018}
{Elmegreen} B.~G.,  2018, \mn@doi [\apj] {10.3847/1538-4357/aaa770}, \href
  {http://adsabs.harvard.edu/abs/2018ApJ...854...16E} {854, 16}

\bibitem[\protect\citeauthoryear{{Elmegreen} \& {Hunter}}{{Elmegreen} \&
  {Hunter}}{2015}]{Elmegreen_Hunter_2015}
{Elmegreen} B.~G.,  {Hunter} D.~A.,  2015, \mn@doi [\apj]
  {10.1088/0004-637X/805/2/145}, \href
  {http://adsabs.harvard.edu/abs/2015ApJ...805..145E} {805, 145}

\bibitem[\protect\citeauthoryear{{Elson}, {de Blok}  \&
  {Kraan-Korteweg}}{{Elson} et~al.}{2012}]{Elson_2012}
{Elson} E.~C.,  {de Blok} W.~J.~G.,   {Kraan-Korteweg} R.~C.,  2012, \mn@doi
  [\aj] {10.1088/0004-6256/143/1/1}, \href
  {http://adsabs.harvard.edu/abs/2012AJ....143....1E} {143, 1}

\bibitem[\protect\citeauthoryear{{Gil de Paz} et~al.,}{{Gil de Paz}
  et~al.}{2007}]{NGS}
{Gil de Paz} A.,  et~al., 2007, \mn@doi [\apjs] {10.1086/516636}, \href
  {http://adsabs.harvard.edu/abs/2007ApJS..173..185G} {173, 185}

\bibitem[\protect\citeauthoryear{{Giovanelli} \& {Haynes}}{{Giovanelli} \&
  {Haynes}}{2002}]{Giovanelli_2002}
{Giovanelli} R.,  {Haynes} M.~P.,  2002, \mn@doi [\apjl] {10.1086/341368},
  \href {http://adsabs.harvard.edu/abs/2002ApJ...571L.107G} {571, L107}

\bibitem[\protect\citeauthoryear{{Gratier} et~al.,}{{Gratier}
  et~al.}{2010}]{Gratier_2010}
{Gratier} P.,  et~al., 2010, \mn@doi [\aap] {10.1051/0004-6361/201014441},
  \href {http://adsabs.harvard.edu/abs/2010A%26A...522A...3G} {522, A3}

\bibitem[\protect\citeauthoryear{{Gratier} et~al.,}{{Gratier}
  et~al.}{2012}]{Gratier_2012}
{Gratier} P.,  et~al., 2012, \mn@doi [\aap] {10.1051/0004-6361/201116612},
  \href {http://adsabs.harvard.edu/abs/2012A%26A...542A.108G} {542, A108}

\bibitem[\protect\citeauthoryear{{Hunter}, {Elmegreen}  \& {Baker}}{{Hunter}
  et~al.}{1998}]{Hunter_1998}
{Hunter} D.~A.,  {Elmegreen} B.~G.,   {Baker} A.~L.,  1998, \mn@doi [\apj]
  {10.1086/305158}, \href {http://adsabs.harvard.edu/abs/1998ApJ...493..595H}
  {493, 595}

\bibitem[\protect\citeauthoryear{{Jarrett} et~al.,}{{Jarrett}
  et~al.}{2012}]{WERGA1}
{Jarrett} T.~H.,  et~al., 2012, \mn@doi [\aj] {10.1088/0004-6256/144/2/68},
  \href {http://adsabs.harvard.edu/abs/2012AJ....144...68J} {144, 68}

\bibitem[\protect\citeauthoryear{{Jarrett} et~al.,}{{Jarrett}
  et~al.}{2013}]{Jarrett_2013}
{Jarrett} T.~H.,  et~al., 2013, \mn@doi [\aj] {10.1088/0004-6256/145/1/6},
  \href {http://adsabs.harvard.edu/abs/2013AJ....145....6J} {145, 6}

\bibitem[\protect\citeauthoryear{{Kam}, {Carignan}, {Chemin}, {Amram}  \&
  {Epinat}}{{Kam} et~al.}{2015}]{Kam_2015}
{Kam} Z.~S.,  {Carignan} C.,  {Chemin} L.,  {Amram} P.,   {Epinat} B.,  2015,
  \mn@doi [\mnras] {10.1093/mnras/stv517}, \href
  {http://adsabs.harvard.edu/abs/2015MNRAS.449.4048K} {449, 4048}

\bibitem[\protect\citeauthoryear{{Kam}, {Carignan}, {Chemin}, {Foster}, {Elson}
   \& {Jarrett}}{{Kam} et~al.}{2017}]{Kam_2017}
{Kam} S.~Z.,  {Carignan} C.,  {Chemin} L.,  {Foster} T.,  {Elson} E.,
  {Jarrett} T.~H.,  2017, \mn@doi [\aj] {10.3847/1538-3881/aa79f3}, \href
  {http://adsabs.harvard.edu/abs/2017AJ....154...41K} {154, 41}

\bibitem[\protect\citeauthoryear{{Kennicutt}}{{Kennicutt}}{1989}]{Kennicutt_1989}
{Kennicutt} Jr. R.~C.,  1989, \mn@doi [\apj] {10.1086/167834}, \href
  {http://adsabs.harvard.edu/abs/1989ApJ...344..685K} {344, 685}

\bibitem[\protect\citeauthoryear{{Kennicutt}}{{Kennicutt}}{1998}]{Kennicutt_1998}
{Kennicutt} Jr. R.~C.,  1998, \mn@doi [\apj] {10.1086/305588}, \href
  {http://adsabs.harvard.edu/abs/1998ApJ...498..541K} {498, 541}

\bibitem[\protect\citeauthoryear{{Kennicutt} \& {Evans}}{{Kennicutt} \&
  {Evans}}{2012}]{Kennicutt_Evans_2012}
{Kennicutt} R.~C.,  {Evans} N.~J.,  2012, \mn@doi [\araa]
  {10.1146/annurev-astro-081811-125610}, \href
  {http://adsabs.harvard.edu/abs/2012ARA%26A..50..531K} {50, 531}

\bibitem[\protect\citeauthoryear{{Kennicutt} Jr. et~al.,}{{Kennicutt}
  et~al.}{2007}]{Kennicutt_2007}
{Kennicutt} Jr. R.~C.,  et~al., 2007, \mn@doi [\apj] {10.1086/522300}, \href
  {http://adsabs.harvard.edu/abs/2007ApJ...671..333K} {671, 333}

\bibitem[\protect\citeauthoryear{{Kennicutt} Jr. et~al.,}{{Kennicutt}
  et~al.}{2009}]{Kennicutt_2009}
{Kennicutt} Jr. R.~C.,  et~al., 2009, \mn@doi [\apj]
  {10.1088/0004-637X/703/2/1672}, \href
  {http://adsabs.harvard.edu/abs/2009ApJ...703.1672K} {703, 1672}

\bibitem[\protect\citeauthoryear{{Kramer} et~al.,}{{Kramer}
  et~al.}{2010}]{HerM33es}
{Kramer} C.,  et~al., 2010, \mn@doi [\aap] {10.1051/0004-6361/201014613}, \href
  {http://adsabs.harvard.edu/abs/2010A%26A...518L..67K} {518, L67}

\bibitem[\protect\citeauthoryear{{Kroupa}}{{Kroupa}}{2001}]{Kroupa_2001}
{Kroupa} P.,  2001, \mn@doi [\mnras] {10.1046/j.1365-8711.2001.04022.x}, \href
  {http://adsabs.harvard.edu/abs/2001MNRAS.322..231K} {322, 231}

\bibitem[\protect\citeauthoryear{{Leroy}, {Walter}, {Brinks}, {Bigiel}, {de
  Blok}, {Madore}  \& {Thornley}}{{Leroy} et~al.}{2008}]{THINGS_Leroy}
{Leroy} A.~K.,  {Walter} F.,  {Brinks} E.,  {Bigiel} F.,  {de Blok} W.~J.~G.,
  {Madore} B.,   {Thornley} M.~D.,  2008, \mn@doi [\aj]
  {10.1088/0004-6256/136/6/2782}, \href
  {http://adsabs.harvard.edu/abs/2008AJ....136.2782L} {136, 2782}

\bibitem[\protect\citeauthoryear{{Martin} \& {Kennicutt}}{{Martin} \&
  {Kennicutt}}{2001}]{Martin_Kennicutt_2001}
{Martin} C.~L.,  {Kennicutt} Jr. R.~C.,  2001, \mn@doi [\apj] {10.1086/321452},
  \href {http://adsabs.harvard.edu/abs/2001ApJ...555..301M} {555, 301}

\bibitem[\protect\citeauthoryear{{Martin} et~al.,}{{Martin}
  et~al.}{2005}]{Martin_2005a}
{Martin} D.~C.,  et~al., 2005, \mn@doi [\apjl] {10.1086/426387}, \href
  {http://adsabs.harvard.edu/abs/2005ApJ...619L...1M} {619, L1}

\bibitem[\protect\citeauthoryear{{Martinsson}}{{Martinsson}}{2011}]{Martinsson_2011}
{Martinsson} T.~P.~K.,  2011, PhD thesis, University of Groningen

\bibitem[\protect\citeauthoryear{{Masci} \& {Fowler}}{{Masci} \&
  {Fowler}}{2009}]{Masci_2009}
{Masci} F.~J.,  {Fowler} J.~W.,  2009, in {Bohlender} D.~A.,  {Durand} D.,
  {Dowler} P.,  eds,  Astronomical Society of the Pacific Conference Series
  Vol. 411, Astronomical Data Analysis Software and Systems XVIII. p.~67
  (\mn@eprint {arXiv} {0812.4310})

\bibitem[\protect\citeauthoryear{{Meurer}, {Zheng}  \& {de Blok}}{{Meurer}
  et~al.}{2013}]{Meurer_2013}
{Meurer} G.~R.,  {Zheng} Z.,   {de Blok} W.~J.~G.,  2013, \mn@doi [\mnras]
  {10.1093/mnras/sts524}, \href
  {http://adsabs.harvard.edu/abs/2013MNRAS.429.2537M} {429, 2537}

\bibitem[\protect\citeauthoryear{{Mogotsi}, {de Blok}, {Cald{\'u}-Primo},
  {Walter}, {Ianjamasimanana}  \& {Leroy}}{{Mogotsi}
  et~al.}{2016}]{Mogotsi_2016}
{Mogotsi} K.~M.,  {de Blok} W.~J.~G.,  {Cald{\'u}-Primo} A.,  {Walter} F.,
  {Ianjamasimanana} R.,   {Leroy} A.~K.,  2016, \mn@doi [\aj]
  {10.3847/0004-6256/151/1/15}, \href
  {http://adsabs.harvard.edu/abs/2016AJ....151...15M} {151, 15}

\bibitem[\protect\citeauthoryear{{Onodera} et~al.,}{{Onodera}
  et~al.}{2010}]{Onodera_2010}
{Onodera} S.,  et~al., 2010, \mn@doi [\apjl] {10.1088/2041-8205/722/2/L127},
  \href {http://adsabs.harvard.edu/abs/2010ApJ...722L.127O} {722, L127}

\bibitem[\protect\citeauthoryear{{P{\'e}rez-Gonz{\'a}lez}
  et~al.,}{{P{\'e}rez-Gonz{\'a}lez} et~al.}{2006}]{Gonzalez_2006}
{P{\'e}rez-Gonz{\'a}lez} P.~G.,  et~al., 2006, \mn@doi [\apj] {10.1086/506196},
  \href {http://adsabs.harvard.edu/abs/2006ApJ...648..987P} {648, 987}

\bibitem[\protect\citeauthoryear{{Rieke} et~al.,}{{Rieke} et~al.}{2004}]{MIPS}
{Rieke} G.~H.,  et~al., 2004, \mn@doi [\apjs] {10.1086/422717}, \href
  {http://adsabs.harvard.edu/abs/2004ApJS..154...25R} {154, 25}

\bibitem[\protect\citeauthoryear{{Rieke}, {Alonso-Herrero}, {Weiner},
  {P{\'e}rez-Gonz{\'a}lez}, {Blaylock}, {Donley}  \& {Marcillac}}{{Rieke}
  et~al.}{2009}]{Rieke_2009}
{Rieke} G.~H.,  {Alonso-Herrero} A.,  {Weiner} B.~J.,  {P{\'e}rez-Gonz{\'a}lez}
  P.~G.,  {Blaylock} M.,  {Donley} J.~L.,   {Marcillac} D.,  2009, \mn@doi
  [\apj] {10.1088/0004-637X/692/1/556}, \href
  {http://adsabs.harvard.edu/abs/2009ApJ...692..556R} {692, 556}

\bibitem[\protect\citeauthoryear{{Rogstad}, {Lockhart}  \& {Wright}}{{Rogstad}
  et~al.}{1974}]{Rogstad_1974}
{Rogstad} D.~H.,  {Lockhart} I.~A.,   {Wright} M.~C.~H.,  1974, \mn@doi [\apj]
  {10.1086/153164}, \href {http://cdsads.u-strasbg.fr/abs/1974ApJ...193..309R}
  {193, 309}

\bibitem[\protect\citeauthoryear{{Salim} et~al.,}{{Salim}
  et~al.}{2007}]{Salim_2007}
{Salim} S.,  et~al., 2007, \mn@doi [\apjs] {10.1086/519218}, \href
  {http://adsabs.harvard.edu/abs/2007ApJS..173..267S} {173, 267}

\bibitem[\protect\citeauthoryear{{Schlegel}, {Finkbeiner}  \&
  {Davis}}{{Schlegel} et~al.}{1998}]{Schlegel_1998}
{Schlegel} D.~J.,  {Finkbeiner} D.~P.,   {Davis} M.,  1998, \mn@doi [\apj]
  {10.1086/305772}, \href {http://adsabs.harvard.edu/abs/1998ApJ...500..525S}
  {500, 525}

\bibitem[\protect\citeauthoryear{{Schruba}, {Leroy}, {Walter}, {Sandstrom}  \&
  {Rosolowsky}}{{Schruba} et~al.}{2010}]{Schruba_2010}
{Schruba} A.,  {Leroy} A.~K.,  {Walter} F.,  {Sandstrom} K.,   {Rosolowsky} E.,
   2010, \mn@doi [\apj] {10.1088/0004-637X/722/2/1699}, \href
  {http://adsabs.harvard.edu/abs/2010ApJ...722.1699S} {722, 1699}

\bibitem[\protect\citeauthoryear{{Schuster} et~al.,}{{Schuster}
  et~al.}{2004}]{Schuster_2004}
{Schuster} K.-F.,  et~al., 2004, \mn@doi [\aap] {10.1051/0004-6361:20034179},
  \href {http://adsabs.harvard.edu/abs/2004A%26A...423.1171S} {423, 1171}

\bibitem[\protect\citeauthoryear{{Tabatabaei} et~al.,}{{Tabatabaei}
  et~al.}{2007}]{Tabatabaei_2007}
{Tabatabaei} F.~S.,  et~al., 2007, \mn@doi [\aap] {10.1051/0004-6361:20066731},
  \href {http://adsabs.harvard.edu/abs/2007A%26A...466..509T} {466, 509}

\bibitem[\protect\citeauthoryear{{Toomre}}{{Toomre}}{1964}]{Toomre_1964}
{Toomre} A.,  1964, \mn@doi [\apj] {10.1086/147861}, \href
  {http://adsabs.harvard.edu/abs/1964ApJ...139.1217T} {139, 1217}

\bibitem[\protect\citeauthoryear{{Williams}, {Gear}  \& {Smith}}{{Williams}
  et~al.}{2018}]{Williams_2018}
{Williams} T.~G.,  {Gear} W.~K.,   {Smith} M.~W.~L.,  2018, \mn@doi [\mnras]
  {10.1093/mnras/sty1476}, \href
  {http://adsabs.harvard.edu/abs/2018MNRAS.479..297W} {479, 297}

\bibitem[\protect\citeauthoryear{{Wong} \& {Blitz}}{{Wong} \&
  {Blitz}}{2002}]{Wong_Blitz_2002}
{Wong} T.,  {Blitz} L.,  2002, \mn@doi [\apj] {10.1086/339287}, \href
  {http://adsabs.harvard.edu/abs/2002ApJ...569..157W} {569, 157}

\bibitem[\protect\citeauthoryear{{Wright} et~al.,}{{Wright}
  et~al.}{2010}]{WISE_Wright}
{Wright} E.~L.,  et~al., 2010, \mn@doi [\aj] {10.1088/0004-6256/140/6/1868},
  \href {http://adsabs.harvard.edu/abs/2010AJ....140.1868W} {140, 1868}

\bibitem[\protect\citeauthoryear{{Wyder} et~al.,}{{Wyder}
  et~al.}{2007}]{Wyder_2007}
{Wyder} T.~K.,  et~al., 2007, \mn@doi [\apjs] {10.1086/521402}, \href
  {http://adsabs.harvard.edu/abs/2007ApJS..173..293W} {173, 293}

\makeatother
\end{thebibliography}




\bsp	
\label{lastpage}
\end{document}